\documentclass[a4paper,11pt]{article}

\usepackage{jheppub} 

\usepackage[T1]{fontenc} 
\usepackage{amsmath}
\usepackage{amssymb}
\usepackage{graphicx}
\usepackage{mathtools}
\usepackage{relsize}
\usepackage{lmodern}
\usepackage{slantsc}
\usepackage{scalefnt}
\usepackage{subfigure}
\usepackage{dsfont}
\usepackage{xspace}
\usepackage{ifpdf}
\usepackage{pgffor}
\usepackage{verbatim}
\usepackage{tikz}
\usetikzlibrary{shapes.geometric}
\usepackage{color}
\usetikzlibrary{arrows,matrix,calc,scopes,decorations.markings}

\allowdisplaybreaks[1]

\newtheorem{theorem}{Theorem}

\newtheorem{convention}[theorem]{Convention}
\newenvironment{proof}[1][Proof]{\noindent\textbf{#1.} }{\ \rule{0.5em}{0.5em}}
\newcommand{\dd}{\mathrm{d}}
\newcommand{\be}{\begin{equation}}
\newcommand{\ee}{\end{equation}}
\newcommand{\bse}{\begin{subequations}}
\newcommand{\ese}{\end{subequations}}
\newcommand{\ket}[1]{|{#1}\rangle}

\newcommand{\braket}[2]{\langle{#1}|{#2}\rangle}

\newcommand{\Z}{\mathbb{Z}}

\newcommand{\ii}{\mathrm{i}}

\newcommand{\Hil}{\mathcal{H}}

\newcommand{\bpm}{\begin{pmatrix}}
\newcommand{\epm}{\end{pmatrix}}
\newcommand{\bmm}{\begin{matrix}}
\newcommand{\emm}{\end{matrix}}

\newcommand{\Blangle}{\Biggl\langle\bmm} 
\newcommand{\BRvert}{\emm\Biggr\vert} 
\newcommand{\BLvert}{\Biggl\vert\bmm} 
\newcommand{\Brangle}{\emm\Biggr\rangle}

\newcommand{\x}{\times}
\newcommand{\fc}{$4$-cocycle\xspace}

\newcommand{\sgn}[1]{\mathrm{sgn}(#1)}
\makeatletter
\newcommand*{\Relbarfill@}{\arrowfill@\Relbar\Relbar\Relbar}
\newcommand*{\xeq}[2][]{\ext@arrow 0055\Relbarfill@{#1}{#2}}
\makeatother

\usepackage{varioref}

\tikzset{->-/.style={decoration={
  markings,
  mark=at position .5 with {\arrow{>}}},postaction={decorate}}}
\tikzset{-<-/.style={decoration={
  markings,
  mark=at position .5 with {\arrow{<}}},postaction={decorate}}}

\newcommand{\smalltorusCube}[9]{
\begin{tikzpicture}[y=0.80pt, x=0.8pt,yscale=-1, inner sep=0pt, outer sep=0pt,scale=#9]
  \begin{scope}[cm={{0.08284,0.0,0.0,0.08284,(38.62164,986.2694)}}]
    \begin{scope}[cm={{1.06299,0.0,0.0,-1.06299,(-186.02362,789.27165)}},draw=black,line join=miter,line cap=butt,miter limit=10.43]
    \end{scope}
  \end{scope}
  \path[draw=black,line join=round,line cap=round,miter limit=4.00,line
    width=0.400pt] (50.2500,1032.1122) -- (50.2500,1002.1122);
  \path[draw=black,line join=round,line cap=round,miter limit=4.00,line
    width=0.400pt] (20.2500,1032.1122) -- (20.2500,1002.1122);
  \path[draw=black,line join=round,line cap=round,miter limit=4.00,line
    width=0.400pt] (50.2500,1002.1122) -- (20.2500,1002.1122);
  \path[draw=black,line join=round,line cap=round,miter limit=4.00,line
    width=0.400pt] (30.2500,992.1122) -- (20.2500,1002.1122);
  \path[draw=black,line join=round,line cap=round,miter limit=4.00,line
    width=0.400pt] (60.2500,992.1122) -- (50.2500,1002.1122);
  \path[draw=black,line join=round,line cap=round,miter limit=4.00,line
    width=0.400pt] (30.2500,992.1122) -- (60.2500,992.1122);
  \path[draw=black,line join=round,line cap=round,miter limit=4.00,line
    width=0.400pt] (60.2500,1022.1122) -- (60.2500,992.1122);
  \path[draw=black,line join=round,line cap=round,miter limit=4.00,line
    width=0.400pt] (50.2500,1032.1122) -- (60.2500,1022.1122);
  \path[draw=black,line join=round,line cap=round,miter limit=4.00,line
    width=0.200pt] (30.2500,992.1122) -- (30.2500,1001.0747)(30.2500,1003.3476) --
    (30.2500,1010.4186)(30.2500,1013.9542) -- (30.2500,1022.1122);
  \path[draw=black,line join=round,line cap=round,miter limit=4.00,line
    width=0.200pt] (60.2500,1022.1122) -- (51.5400,1022.1122)(49.0146,1022.1122)
    -- (41.4385,1022.1122)(38.6605,1022.1122) -- (30.2500,1022.1122);
  \path[draw=black,line join=round,line cap=round,miter limit=4.00,line
    width=0.200pt] (20.2500,1032.1122) -- (30.2500,1022.1122);
  \path[draw=black,line join=round,line cap=round,miter limit=4.00,line
    width=0.400pt] (20.2500,1002.1122) -- (50.2500,1032.1122);
  \path[draw=black,line join=round,line cap=round,miter limit=4.00,line
    width=0.400pt] (50.2500,1002.1122) -- (60.2500,1022.1122);
  \path[draw=black,line join=round,line cap=round,miter limit=4.00,line
    width=0.400pt] (30.2500,992.1122) -- (50.2500,1002.1122);
  \path[draw=black,line join=round,line cap=round,miter limit=4.00,line
    width=0.200pt] (20.2500,1002.1122) -- (30.2500,1022.1122);
  \path[draw=black,line join=round,line cap=round,miter limit=4.00,line
    width=0.200pt] (30.2500,1022.1122) -- (50.2500,1032.1122);
  \path[draw=black,line join=round,line cap=round,miter limit=4.00,line
    width=0.200pt] (30.2500,992.1122) -- (39.4416,1001.3038)(41.2094,1003.0716) --
    (44.4292,1006.2915)(46.1970,1008.0592) --
    (49.2275,1011.0897)(51.3109,1013.1731) -- (60.2500,1022.1122);
  \path[draw=black,line join=round,line cap=round,miter limit=4.00,line
    width=0.200pt] (30.2500,1022.1122) -- (34.4305,1017.9317)(36.0089,1016.3533)
    -- (50.2500,1002.1122);
  \path[fill=black] (16.399113,1036.236) node[above right]  {\scalebox{0.5}{$#1$}};
  \path[fill=black] (53.447334,1036.1368) node[above right] {\scalebox{0.5}{$#2$}};
  \path[fill=black] (24.25,1022.1121) node[above right] {\scalebox{0.5}{$#3$}};
  \path[fill=black] (61.492386,1022.1121) node[above right]  {\scalebox{0.5}{$#4$}};
  \path[fill=black] (16.328045,1003.2214) node[above right]  {\scalebox{0.5}{$#5$}};
  \path[fill=black] (54.366116,1002.9758) node[above right] {\scalebox{0.5}{$#6$}};
  \path[fill=black] (26.891497,991.84943) node[above right]{\scalebox{0.5}{$#7$}};
  \path[fill=black] (61.156353,991.97632) node[above right] {\scalebox{0.5}{$#8$}};
  \path[draw=black,line join=round,line cap=round,miter limit=4.00,line
    width=0.400pt] (20.2500,1032.1122) -- (50.2500,1032.1122);
  \path[fill=black] (61.4,1010) node[above right] {\scalebox{0.65}{$k$}};
  \path[fill=black] (44,1021) node[above right] {\scalebox{0.65}{$g$}};
  \path[fill=black] (57,1031.0608) node[above right] {\scalebox{0.65}{$h$}};
  \path[fill=black] (44,990) node[above right] {\scalebox{0.65}{$g^{'}$}};
   \path[fill=black] (20,998.0608) node[above right] {\scalebox{0.65}{$h'$}};
 \path [gray,fill, opacity=0.4, opacity=0.3] (20.2500,1032.1122) -- (50.2500,1032.1122) -- (60.2500,1022.1122) -- (30.2500,1022.1122)--cycle;
\path [lightgray,fill, opacity=0.4, opacity=0.3] (20.2500,1002.1122) -- (50.2500,1002.1122) -- (60.2500,992.1122) -- (30.2500,992.1122)--cycle;
\end{tikzpicture}
}
\newcommand{\Avprojector}{
\begin{tikzpicture}[scale=0.5]
\draw [fill] (5.6,5.2) -- (3.127,5.695);
\draw [fill] (2.931,5.734) -- (1.6,6.);
\draw [fill] (5.6,5.2) -- (1.6,4.);
\draw [fill] (1.6,6.) -- (1.6,4.);
\draw [fill] (5.6,5.2) -- (5.6,8.8);
\draw [fill] (1.6,4.) -- (5.6,8.8);
\draw [fill] (1.6,6.) -- (5.6,8.8);
\draw [fill] (5.6,5.2) -- (7.493,6.147);
\draw [fill] (7.672,6.236) -- (8.444,6.622);
\draw [fill] (8.623,6.711) -- (9.2,7.);
\draw [fill] (5.6,5.2) -- (8.06,5.424);
\draw [fill] (8.259,5.442) -- (10.,5.6);
\draw [fill] (5.6,5.2) -- (9.4,3.8);
\draw [fill] (9.2,7.) -- (10.,5.6);
\draw [fill] (10.,5.6) -- (9.4,3.8);
\draw [fill] (5.6,8.8) -- (9.4,3.8);
\draw [fill] (5.6,8.8) -- (10.,5.6);
\draw [fill] (5.6,8.8) -- (9.2,7.);
\draw[dashed] (1.6,4.) -- (0.8,3.4);
\draw [dashed] (1.6,4.) -- (2.4,3.4);
\draw [dashed] (1.6,6.) -- (0.4,7.8);
\draw [dashed] (1.6,6.) -- (2.6,8.2);
\draw [dashed] (9.2,7.) -- (10.,8.6);
\draw [dashed] (9.2,7.) -- (11.2,6.3);
\draw [dashed] (10.,5.6) -- (11.2,4.6);
\draw [dashed] (9.4,3.8) -- (9.6,3.2);
\draw [dashed] (9.4,3.8) -- (11.,4.);
\node[above] at(5.6,8.8) {0};
\node[left] at(1.6,4) {$v_N$};
\node[left] at(1.6,6) {$v_1$};
\node[below] at (5.6,5.2) {1};
\node[left] at (9.4,3.8){$v_{i+2}$};
\node[right] at (10.,5.6){$v_{i+1}$};
\node[right] at (9.2,7.){$v_i$};
\end{tikzpicture}
}
\newcommand{\Avcommutator}[7]{
\begin{tikzpicture}[scale=#7,baseline]
\coordinate (a) at(0,0);    
\coordinate (b) at (1,2);  
\coordinate (c) at (2.5,0);    
\coordinate (d) at (3.5,2);     
\coordinate (n) at (2,3); 
\draw (a) -- (b) -- (n) --  cycle;
\draw (a)  -- (c) -- (n) -- cycle;
\draw (n) -- (c)-- (d)-- cycle;
\draw (b)--(1.2,2);\draw (1.5,2) --(2.1,2);
\draw (2.3,2) --(d);
\draw (a) --(2.2,1.25);
\draw (2.4,1.35) --(d);

\draw[dashed] (a)--(-0.6,-0.6);
\draw[dashed] (b)--(0.6,2.7);
\draw[dashed] (c)--(3,-0.6);
\draw[dashed] (d)--(4,2.7);

\node[left] at(a) {\scalebox{0.7}{$#1$}};
\node[above] at(b) {\scalebox{0.7}{$#2$}};
\node[right] at(c) {\scalebox{0.7}{$#4$}};
\node[above] at(d) {\scalebox{0.7}{$#3$}};
\node[right] at(n) {\scalebox{0.7}{$#6$}};
\ifnum #5>0 {
\node[circle,fill=black,outer sep=0pt, inner sep=0.6pt] at(n) {};
\draw (a) -- (n);
\draw (b) -- (n);
\draw (c) -- (n);
\draw (d) -- (n);
}\fi

\end{tikzpicture}
}
\newcommand{\Avamplitude}[7]{
\begin{tikzpicture}[scale=#7,baseline]
\coordinate (a) at(0,0);    
\coordinate (b) at (0.8,-0.75);  
\coordinate (c) at (2.5,0);      
\coordinate (d) at (1.25,2);     
\coordinate (e) at ($ (a) ! 0.3 ! (c) $);
\coordinate (f) at ($ (a) ! 0.45 ! (c) $);
\coordinate (g) at ($ (a) ! 0.5 ! (b) $);
\coordinate (h) at ($ (c) ! 0.5 ! (d) $);
\coordinate (m) at (-0.7,0.6);
\coordinate (k) at (3.5,0.5);
\coordinate (l) at (0.1,-1.5);
\coordinate (n) at (1.2,-1.8); 
\draw (a) -- (b) -- (c) -- (d) -- cycle;
\draw (b) -- (d);
\draw (a) -- (e);
\draw (f) -- (c);
\path [gray, fill, opacity=0.5, opacity=0.3] (a) -- (n) -- (b) -- (d)--cycle;
\path [gray, fill, opacity=0.5, opacity=0.3] (d) -- (b) -- (n) --(c)-- cycle;

\node[left] at(a) {\scalebox{0.7}{$#1$}};
\node[right] at(b) {\scalebox{0.7}{$#2$}};
\node[right] at(c) {\scalebox{0.7}{$#4$}};
\node[above] at(d) {\scalebox{0.7}{$#3$}};
\node[right] at(n) {\scalebox{0.7}{$#6$}};
\ifnum #5>0 {
\node[circle,fill=black,outer sep=0pt, inner sep=0.1pt] at(n) {};
\draw (a) -- (n);
\draw (b) -- (n);
\draw (c) -- (n);
\draw[dashed] (d) -- (n);
\draw[dashed] (a) -- (m);
\draw[dashed] (b) -- (l);
\draw[dashed] (c) -- (k);
}\fi

\end{tikzpicture}
}
\newcommand{\frobenius}[8]{
\begin{tikzpicture}[scale=#7,baseline]
\coordinate (a) at(0,0);    
\coordinate (b) at (0.8,-0.75);  
\coordinate (c) at (2.5,0);      
\coordinate (d) at (1.25,2);     
\coordinate (e) at ($ (a) ! 0.3 ! (c) $);
\coordinate (f) at ($ (a) ! 0.45 ! (c) $);
\coordinate (g) at ($ (a) ! 0.5 ! (b) $);
\coordinate (h) at ($ (c) ! 0.5 ! (d) $);
\coordinate (n) at ($ (g) ! 0.5 ! (h) $); 
\draw (a) -- (b) -- (c) -- (d) -- cycle;
\draw (b) -- (d);
\draw (a) -- (e);
\draw (f) -- (c);
\path [gray, fill, opacity=0.3] (a) -- (b) -- (d) -- cycle;
\path [gray, fill, opacity=0.3] (d) -- (b) -- (c) -- cycle;
\node[left] at(a) {\scalebox{0.7}{$#1$}};
\node[below] at(b) {\scalebox{0.7}{$#2$}};
\node[right] at(c) {\scalebox{0.7}{$#4$}};
\node[above] at(d) {\scalebox{0.7}{$#3$}};
\node[right] at(n) {\scalebox{0.7}{$#6$}};
\ifnum #5>0 {
\node[circle,fill=black,outer sep=0pt, inner sep=0.6pt] at(n) {};
\draw[dashed] (a) -- (n);
\draw[dashed] (b) -- (n);
\draw[dashed] (c) -- (n);
\draw[dashed] (d) -- (n);
}\fi

\end{tikzpicture}
}
\newcommand{\fourTetboundary}[8]{
\begin{tikzpicture}[scale=#7,baseline]
\coordinate (a) at(0,0);    
\coordinate (b) at (0.8,-0.75);  
\coordinate (c) at (2.5,0);      
\coordinate (d) at (1.25,2);     
\coordinate (e) at ($ (a) ! 0.3 ! (c) $);
\coordinate (f) at ($ (a) ! 0.45 ! (c) $);
\coordinate (g) at ($ (a) ! 0.5 ! (b) $);
\coordinate (h) at ($ (c) ! 0.5 ! (d) $);
\coordinate (m) at (-0.7,0.6);
\coordinate (k) at (3.5,0.5);
\coordinate (l) at (0.1,-1.5);
\coordinate (n) at (1.2,-1.8); 
\draw (a) -- (b) -- (c) -- (d) -- cycle;
\draw (b) -- (d);
\draw (a) -- (e);
\draw (f) -- (c);
\ifnum #8=1 {
\draw[-<,>=latex,line width=0.01pt] (a) -- ($ (a) ! 0.5 ! (b) $);
\draw[-<,>=latex,line width=0.01pt] (b) -- ($ (b) ! 0.5 ! (c) $);
\draw[-<,>=latex,line width=0.01pt] (c) -- ($ (c) ! 0.5 ! (d) $);
\draw[<-,>=latex,line width=0.01pt] (f) -- (c);
\draw[-<,>=latex,line width=0.01pt] (b) -- ($ (b) ! 0.5 ! (d) $);
\draw[-<,>=latex,line width=0.01pt] (a) -- ($ (a) ! 0.5 ! (d) $);}
\fi

\node[left] at(a) {\scalebox{0.7}{$#1$}};
\node[right] at(b) {\scalebox{0.7}{$#2$}};
\node[right] at(c) {\scalebox{0.7}{$#4$}};
\node[above] at(d) {\scalebox{0.7}{$#3$}};
\ifnum #5>0 {
\node[circle,fill=black,outer sep=0pt, inner sep=0.1pt] at(n) {};
\draw[dotted] (a) -- (n);
\draw[dotted] (b) -- (n);
\draw[dotted] (c) -- (n);
}\fi
\draw[dashed] (a) -- (m);
\draw[dashed] (b) -- (l);
\draw[dashed] (c) -- (k);

\end{tikzpicture}
}

\newcommand{\tetrahedronSplit}[9]{
\begin{tikzpicture}[scale=1,baseline]
\coordinate (a) at(0,0);
\coordinate (b) at (0.8,-0.75);
\coordinate (c) at (2.5,0);
\coordinate (d) at (1.25,2);
\coordinate (e) at ($ (a) ! 0.3 ! (c) $);
\coordinate (f) at ($ (a) ! 0.45 ! (c) $);
\coordinate (g) at ($ (a) ! 0.5 ! (b) $);
\coordinate (h) at ($ (c) ! 0.5 ! (d) $);
\coordinate (n) at ($ (g) ! 0.5 ! (h) $); 
\draw (a) -- (b) -- (c) -- (d) -- cycle;
\draw (b) -- (d);
\draw (a) -- (e);
\draw (f) -- (c);
\draw[-<,>=latex,line width=0.01pt] (a) -- ($ (a) ! 0.5 ! (b) $);
\draw[-<,>=latex,line width=0.01pt] (b) -- ($ (b) ! 0.5 ! (c) $);
\draw[-<,>=latex,line width=0.01pt] (c) -- ($ (c) ! 0.5 ! (d) $);
\draw[<-,>=latex,line width=0.01pt] (f) -- (c);
\draw[-<,>=latex,line width=0.01pt] (b) -- ($ (b) ! 0.5 ! (d) $);
\draw[-<,>=latex,line width=0.01pt] (a) -- ($ (a) ! 0.5 ! (d) $);

\node[left] at(a) {\scalebox{0.7}{$#1$}};
\node[below] at(b) {\scalebox{0.7}{$#2$}};
\node[right] at(c) {\scalebox{0.7}{$#3$}};
\node[above] at(d) {\scalebox{0.7}{$#4$}};
\node[circle,fill=black,outer sep=0pt, inner sep=0.6pt] at(n) {};
\node[below] at ($ (a) ! 0.5 ! (b) $) {\scalebox{0.7}{$#5$}};
\node[below] at ($ (b) ! 0.5 ! (c) $) {\scalebox{0.7}{$#6$}};
\node[right] at ($ (c) ! 0.5 ! (d) $) {\scalebox{0.7}{$#7$}};
\node at (0.6,-0.15) {\scalebox{0.7}{$#5#6$}};
\node[left] at ($ (b) ! 0.5 ! (d) $) {\scalebox{0.7}{$#6#7$}};
\node[left] at ($ (a) ! 0.5 ! (d) $) {\scalebox{0.7}{$#5#6#7$}};

\draw[dashed] (a) -- (n);
\draw[dashed] (b) -- (n);
\draw[dashed] (c) -- (n);
\draw[dashed] (d) -- (n);

\ifnum #8=1 {
  \node[right, outer sep=0pt, inner sep=1pt] at($ (n) ! 0.07 ! (d) $) {\scalebox{0.7}{$#1'$}};
  \draw[->,>=latex,line width=0.01pt] ($ (a) ! 0.4 ! (n) $) -- ($ (a) ! 0.5 ! (n) $) node[above] {\scalebox{0.7}{$#9$}};
  \draw[->,>=latex,line width=0.01pt] ($ (b) ! 0.4 ! (n) $) node[right] {\scalebox{0.7}{$#9#5$}} -- ($ (b) ! 0.5 ! (n) $) ;
  \draw[->,>=latex,line width=0.01pt] ($ (c) ! 0.4 ! (n) $) -- ($ (c) ! 0.5 ! (n) $) node[above] {\scalebox{0.7}{$#9#5#6$}};
  \draw[->,>=latex,line width=0.01pt] ($ (d) ! 0.4 ! (n) $) -- ($ (d) ! 0.5 ! (n) $) node[right,outer sep=0pt, inner sep=1pt] {\scalebox{0.6}{$#9#5#6#7$}};}
\fi
\ifnum #8=2 {
  \node[right, outer sep=0pt, inner sep=1pt] at($ (n) ! 0.07 ! (d) $) {\scalebox{0.7}{$#2'$}};
  \draw[-<,>=latex,line width=0.01pt] ($ (a) ! 0.4 ! (n) $) -- ($ (a) ! 0.5 ! (n) $) node[above, outer sep=0pt, inner sep=1pt] {\scalebox{0.7}{$#5#9^{-1}$}};
  \draw[->,>=latex,line width=0.01pt] ($ (b) ! 0.4 ! (n) $) node[right] {\scalebox{0.7}{$#9$}} -- ($ (b) ! 0.5 ! (n) $) ;
  \draw[->,>=latex,line width=0.01pt] ($ (c) ! 0.4 ! (n) $) -- ($ (c) ! 0.5 ! (n) $) node[above] {\scalebox{0.7}{$#9#6$}};
  \draw[->,>=latex,line width=0.01pt] ($ (d) ! 0.4 ! (n) $) -- ($ (d) ! 0.5 ! (n) $) node[right,outer sep=0pt, inner sep=1pt] {\scalebox{0.6}{$#9#6#7$}};}
\fi
\ifnum #8=3 {
  \node[right, outer sep=0pt, inner sep=1pt] at($ (n) ! 0.07 ! (d) $) {\scalebox{0.7}{$#3'$}};
  \draw[-<,>=latex,line width=0.01pt] ($ (a) ! 0.4 ! (n) $) -- ($ (a) ! 0.5 ! (n) $) node[above, outer sep=0pt, inner sep=1pt] {\scalebox{0.7}{$#5#6#9^{-1}$}};
  \draw[-<,>=latex,line width=0.01pt] ($ (b) ! 0.4 ! (n) $) node[right, outer sep=0pt, inner sep=1pt] {\scalebox{0.7}{$#6#9^{-1}$}} -- ($ (b) ! 0.5 ! (n) $) ;
  \draw[->,>=latex,line width=0.01pt] ($ (c) ! 0.4 ! (n) $) -- ($ (c) ! 0.5 ! (n) $) node[above] {\scalebox{0.7}{$#9$}};
  \draw[->,>=latex,line width=0.01pt] ($ (d) ! 0.4 ! (n) $) -- ($ (d) ! 0.5 ! (n) $) node[right,outer sep=0pt, inner sep=1pt] {\scalebox{0.6}{$#9#7$}};}
\fi
\ifnum #8=4 {
  \node[right, outer sep=0pt, inner sep=1pt] at($ (n) ! 0.07 ! (d) $) {\scalebox{0.7}{$#4'$}};
  \draw[-<,>=latex,line width=0.01pt] ($ (a) ! 0.4 ! (n) $) -- ($ (a) ! 0.5 ! (n) $) node[above, outer sep=0pt, inner sep=1pt] {\scalebox{0.7}{$#5#6#7#9^{-1}$}};
  \draw[-<,>=latex,line width=0.01pt] ($ (b) ! 0.4 ! (n) $) node[right, outer sep=0pt, inner sep=1pt] {\scalebox{0.7}{$#6#7#9^{-1}$}} -- ($ (b) ! 0.5 ! (n) $) ;
  \draw[-<,>=latex,line width=0.01pt] ($ (c) ! 0.4 ! (n) $) -- ($ (c) ! 0.5 ! (n) $) node[above] {\scalebox{0.7}{$#7#9^{-1}$}};
  \draw[->,>=latex,line width=0.01pt] ($ (d) ! 0.4 ! (n) $) -- ($ (d) ! 0.5 ! (n) $) node[right,outer sep=0pt, inner sep=1pt] {\scalebox{0.6}{$#9$}};}
\fi
\end{tikzpicture}
}

\newcommand{\oneTriangle}[4][0]{
\begin{tikzpicture}[scale=1,baseline]
\draw (0,0) node[left] {\scalebox{0.7}{$#2$}} -- (1,0) node[right] {\scalebox{0.7}{$#3$}} -- (60:1) node[above] {\scalebox{0.7}{$#4$}} -- cycle;
\ifnum #1=1 {
   \draw[-<,>=latex,line width=0.01pt] (0,0) -- (0.5,0) node[below] {\scalebox{0.7}{$[#2#3]$}};
   \draw[-<,>=latex,line width=0.01pt] (0,0) -- (60:0.5) node[left] {\scalebox{0.7}{$[#3#4]$}};
   \draw[->,>=latex,line width=0.01pt] (0,0) +(60:1) -- +(30:0.866) node[right] {\scalebox{0.7}{$[#2#4]$}};}
\else {
   \draw[-<,>=latex,line width=0.01pt] (0,0) -- (0.5,0) ;
   \draw[-<,>=latex,line width=0.01pt] (0,0) -- (60:0.5) ;
   \draw[->,>=latex,line width=0.01pt] (0,0) +(60:1) -- +(30:0.866);}  \fi
\end{tikzpicture}
}

\newcommand{\twoTriangles}[5]{
\begin{tikzpicture}[scale=0.8,baseline]
\coordinate (b) at (0,0);
\coordinate (c) at (30:1);
\coordinate (d) at (0,1);
\coordinate (a) at (150:1);

\draw (a) node[left] {\scalebox{0.7}{$#1$}} -- (b) node[below] {\scalebox{0.7}{$#2$}} -- (c) node[right] {\scalebox{0.7}{$#3$}} -- (d) node[above] {\scalebox{0.7}{$#4$}} -- cycle;
\draw[-<,>=latex,line width=0.01pt] (a) -- ($ (a) ! 0.5 ! (b) $);
\draw[-<,>=latex,line width=0.01pt] (b) -- ($ (b) ! 0.5 ! (c) $);
\draw[-<,>=latex,line width=0.01pt] (c) -- ($ (c) ! 0.5 ! (d) $);
\draw[-<,>=latex,line width=0.01pt] (a) -- ($ (a) ! 0.5 ! (d) $);

\ifnum #5=0 {
   \draw (b)  -- (d);
   \draw[-<,>=latex,line width=0.01pt] (b) -- ($ (b) ! 0.5 ! (d) $);}
\else {
   \draw (a)  -- (c);
   \draw[-<,>=latex,line width=0.01pt] (a) -- ($ (a) ! 0.5 ! (c) $);}
\fi
\end{tikzpicture}
}

\newcommand{\threeTriangles}[5]{
\begin{tikzpicture}[scale=0.8,baseline]
\coordinate (c) at (0,0);
\coordinate (d) at (0,1);
\coordinate (a) at (210:1);
\coordinate (b) at (-30:1);

\draw (a) node[left] {\scalebox{0.7}{$#1$}} -- (b) node[right] {\scalebox{0.7}{$#2$}} -- (c) node[below] {\scalebox{0.7}{$#3$}} -- (d) node[above] {\scalebox{0.7}{$#4$}} -- (a) -- (c);
\draw (b) -- (d);
\draw[-<,>=latex,line width=0.01pt] (a) -- ($ (a) ! 0.5 ! (b) $);
\draw[-<,>=latex,line width=0.01pt] (a) -- ($ (a) ! 0.5 ! (d) $);
\draw[-<,>=latex,line width=0.01pt] (b) -- ($ (b) ! 0.5 ! (d) $);

\ifnum #5=1 {
\draw[-<,>=latex,line width=0.01pt] (b) -- ($ (b) ! 0.5 ! (c) $);
\draw[->,>=latex,line width=0.01pt] (c) -- ($ (c) ! 0.5 ! (d) $);
\draw[-<,>=latex,line width=0.01pt] (a) -- ($ (a) ! 0.5 ! (c) $);}
\fi
\ifnum #5=2 {
\draw[-<,>=latex,line width=0.01pt] (b) -- ($ (b) ! 0.5 ! (c) $);
\draw[-<,>=latex,line width=0.01pt] (c) -- ($ (c) ! 0.5 ! (d) $);
\draw[-<,>=latex,line width=0.01pt] (a) -- ($ (a) ! 0.5 ! (c) $);}
\fi
\ifnum #5=3 {
\draw[->,>=latex,line width=0.01pt] (b) -- ($ (b) ! 0.5 ! (c) $);
\draw[-<,>=latex,line width=0.01pt] (c) -- ($ (c) ! 0.5 ! (d) $);
\draw[-<,>=latex,line width=0.01pt] (a) -- ($ (a) ! 0.5 ! (c) $);}
\fi
\ifnum #5=4 {
\draw[->,>=latex,line width=0.01pt] (b) -- ($ (b) ! 0.5 ! (c) $);
\draw[-<,>=latex,line width=0.01pt] (c) -- ($ (c) ! 0.5 ! (d) $);
\draw[->,>=latex,line width=0.01pt] (a) -- ($ (a) ! 0.5 ! (c) $);}
\fi
\end{tikzpicture}
}
\newcommand{\fourTet}[8]{
\begin{tikzpicture}[scale=#7,baseline]
\coordinate (a) at(0,0);    
\coordinate (b) at (0.8,-0.75);  
\coordinate (c) at (2.5,0);      
\coordinate (d) at (1.25,2);     
\coordinate (e) at ($ (a) ! 0.3 ! (c) $);
\coordinate (f) at ($ (a) ! 0.45 ! (c) $);
\coordinate (g) at ($ (a) ! 0.5 ! (b) $);
\coordinate (h) at ($ (c) ! 0.5 ! (d) $);
\coordinate (n) at ($ (g) ! 0.5 ! (h) $); 
\draw (a) -- (b) -- (c) -- (d) -- cycle;
\draw (b) -- (d);
\draw (a) -- (e);
\draw (f) -- (c);
\ifnum #8=1 {
\draw[-<,>=latex,line width=0.01pt] (a) -- ($ (a) ! 0.5 ! (b) $);
\draw[-<,>=latex,line width=0.01pt] (b) -- ($ (b) ! 0.5 ! (c) $);
\draw[-<,>=latex,line width=0.01pt] (c) -- ($ (c) ! 0.5 ! (d) $);
\draw[<-,>=latex,line width=0.01pt] (f) -- (c);
\draw[-<,>=latex,line width=0.01pt] (b) -- ($ (b) ! 0.5 ! (d) $);
\draw[-<,>=latex,line width=0.01pt] (a) -- ($ (a) ! 0.5 ! (d) $);}
\fi

\node[left] at(a) {\scalebox{0.7}{$#1$}};
\node[below] at(b) {\scalebox{0.7}{$#2$}};
\node[right] at(c) {\scalebox{0.7}{$#4$}};
\node[above] at(d) {\scalebox{0.7}{$#3$}};
\ifnum #5>0 {
\node[circle,fill=black,outer sep=0pt, inner sep=0.6pt] at(n) {};
\draw[dashed] (a) -- (n);
\draw[dashed] (b) -- (n);
\draw[dashed] (c) -- (n);
\draw[dashed] (d) -- (n);
}\fi

\ifnum #5=1 {
  \node[right, outer sep=0pt, inner sep=1pt] at($ (n) ! 0.07 ! (d) $) {\scalebox{0.7}{$#6$}};
  \ifnum #8=1 {
  \draw[->,>=latex,line width=0.01pt] ($ (a) ! 0.4 ! (n) $) -- ($ (a) ! 0.5 ! (n) $) ;
  \draw[->,>=latex,line width=0.01pt] ($ (b) ! 0.4 ! (n) $) -- ($ (b) ! 0.5 ! (n) $) ;
  \draw[->,>=latex,line width=0.01pt] ($ (c) ! 0.4 ! (n) $) -- ($ (c) ! 0.5 ! (n) $) ;
  \draw[->,>=latex,line width=0.01pt] ($ (d) ! 0.4 ! (n) $) -- ($ (d) ! 0.5 ! (n) $) ;}\fi}
\fi
\ifnum #5=2 {
  \node[right, outer sep=0pt, inner sep=1pt] at($ (n) ! 0.07 ! (d) $) {\scalebox{0.7}{$#6$}};
  \ifnum #8=1 {
  \draw[-<,>=latex,line width=0.01pt] ($ (a) ! 0.4 ! (n) $) -- ($ (a) ! 0.5 ! (n) $) ;
  \draw[->,>=latex,line width=0.01pt] ($ (b) ! 0.4 ! (n) $) -- ($ (b) ! 0.5 ! (n) $) ;
  \draw[->,>=latex,line width=0.01pt] ($ (c) ! 0.4 ! (n) $) -- ($ (c) ! 0.5 ! (n) $) ;
  \draw[->,>=latex,line width=0.01pt] ($ (d) ! 0.4 ! (n) $) -- ($ (d) ! 0.5 ! (n) $) ;}\fi}
\fi
\ifnum #5=3 {
  \node[right, outer sep=0pt, inner sep=1pt] at($ (n) ! 0.07 ! (d) $) {\scalebox{0.7}{$#6$}};
  \ifnum #8=1 {
  \draw[-<,>=latex,line width=0.01pt] ($ (a) ! 0.4 ! (n) $) -- ($ (a) ! 0.5 ! (n) $) ;
  \draw[-<,>=latex,line width=0.01pt] ($ (b) ! 0.4 ! (n) $) -- ($ (b) ! 0.5 ! (n) $) ;
  \draw[->,>=latex,line width=0.01pt] ($ (c) ! 0.4 ! (n) $) -- ($ (c) ! 0.5 ! (n) $) ;
  \draw[->,>=latex,line width=0.01pt] ($ (d) ! 0.4 ! (n) $) -- ($ (d) ! 0.5 ! (n) $) ;}\fi}
\fi
\ifnum #5=4 {
  \node[right, outer sep=0pt, inner sep=1pt] at($ (n) ! 0.07 ! (d) $) {\scalebox{0.7}{$#6$}};
  \ifnum #8=1 {
  \draw[-<,>=latex,line width=0.01pt] ($ (a) ! 0.4 ! (n) $) -- ($ (a) ! 0.5 ! (n) $) ;
  \draw[-<,>=latex,line width=0.01pt] ($ (b) ! 0.4 ! (n) $) -- ($ (b) ! 0.5 ! (n) $) ;
  \draw[-<,>=latex,line width=0.01pt] ($ (c) ! 0.4 ! (n) $) -- ($ (c) ! 0.5 ! (n) $) ;
  \draw[->,>=latex,line width=0.01pt] ($ (d) ! 0.4 ! (n) $) -- ($ (d) ! 0.5 ! (n) $) ;}\fi}
\fi
\end{tikzpicture}
}

\newcommand{\BvTri}[5]{
\begin{tikzpicture}[scale=1,baseline]
\coordinate (a) at(0,0);    
\coordinate (b) at (0.8,-0.75);  
\coordinate (c) at (2.5,0);      
\coordinate (d) at (1.25,2);     
\coordinate (e) at ($ (a) ! 0.3 ! (c) $);
\coordinate (f) at ($ (a) ! 0.45 ! (c) $);
\coordinate (g) at ($ (a) ! 0.5 ! (b) $);
\coordinate (h) at ($ (c) ! 0.5 ! (d) $);
\coordinate (n) at ($ (g) ! 0.5 ! (h) $); 
\draw (a) -- (b) -- (c) -- (d) -- cycle;
\draw (b) -- (d);
\draw (a) -- (e);
\draw (f) -- (c);
\draw[-<,>=latex,line width=0.01pt] (a) -- ($ (a) ! 0.5 ! (b) $);
\draw[-<,>=latex,line width=0.01pt] (b) -- ($ (b) ! 0.5 ! (c) $);
\draw[-<,>=latex,line width=0.01pt] (c) -- ($ (c) ! 0.5 ! (d) $);
\draw[<-,>=latex,line width=0.01pt] (f) -- (c);
\draw[-<,>=latex,line width=0.01pt] (b) -- ($ (b) ! 0.5 ! (d) $);
\draw[-<,>=latex,line width=0.01pt] (a) -- ($ (a) ! 0.5 ! (d) $);

\node[left] at(a) {\scalebox{0.7}{$#1$}};
\node[below] at(b) {\scalebox{0.7}{$#2$}};
\node[right] at(c) {\scalebox{0.7}{$#4$}};
\node[above] at(d) {\scalebox{0.7}{$#3$}};
\node[circle,fill=black,outer sep=0pt, inner sep=0.6pt] at(n) {};
\draw[dashed] (a) -- (n);
\draw[dashed] (b) -- (n);
\draw[dashed] (c) -- (n);
\draw[dashed] (d) -- (n);

\ifnum #5=1 {
  \node[right, outer sep=0pt, inner sep=1pt] at($ (n) ! 0.07 ! (d) $) {\scalebox{0.7}{$#1'$}};
  \draw[->,>=latex,line width=0.01pt] ($ (a) ! 0.4 ! (n) $) -- ($ (a) ! 0.5 ! (n) $) ;
  \draw[->,>=latex,line width=0.01pt] ($ (b) ! 0.4 ! (n) $) -- ($ (b) ! 0.5 ! (n) $) ;
  \draw[->,>=latex,line width=0.01pt] ($ (c) ! 0.4 ! (n) $) -- ($ (c) ! 0.5 ! (n) $) ;
  \draw[->,>=latex,line width=0.01pt] ($ (d) ! 0.4 ! (n) $) -- ($ (d) ! 0.5 ! (n) $) ;}
\fi
\ifnum #5=2 {
  \node[right, outer sep=0pt, inner sep=1pt] at($ (n) ! 0.07 ! (d) $) {\scalebox{0.7}{$#2'$}};
  \draw[-<,>=latex,line width=0.01pt] ($ (a) ! 0.4 ! (n) $) -- ($ (a) ! 0.5 ! (n) $) ;
  \draw[->,>=latex,line width=0.01pt] ($ (b) ! 0.4 ! (n) $) -- ($ (b) ! 0.5 ! (n) $) ;
  \draw[->,>=latex,line width=0.01pt] ($ (c) ! 0.4 ! (n) $) -- ($ (c) ! 0.5 ! (n) $) ;
  \draw[->,>=latex,line width=0.01pt] ($ (d) ! 0.4 ! (n) $) -- ($ (d) ! 0.5 ! (n) $) ;}
\fi
\ifnum #5=3 {
  \node[right, outer sep=0pt, inner sep=1pt] at($ (n) ! 0.07 ! (d) $) {\scalebox{0.7}{$#3'$}};
  \draw[-<,>=latex,line width=0.01pt] ($ (a) ! 0.4 ! (n) $) -- ($ (a) ! 0.5 ! (n) $) ;
  \draw[-<,>=latex,line width=0.01pt] ($ (b) ! 0.4 ! (n) $) -- ($ (b) ! 0.5 ! (n) $) ;
  \draw[->,>=latex,line width=0.01pt] ($ (c) ! 0.4 ! (n) $) -- ($ (c) ! 0.5 ! (n) $) ;
  \draw[->,>=latex,line width=0.01pt] ($ (d) ! 0.4 ! (n) $) -- ($ (d) ! 0.5 ! (n) $) ;}
\fi
\ifnum #5=4 {
  \node[right, outer sep=0pt, inner sep=1pt] at($ (n) ! 0.07 ! (d) $) {\scalebox{0.7}{$#4'$}};
  \draw[-<,>=latex,line width=0.01pt] ($ (a) ! 0.4 ! (n) $) -- ($ (a) ! 0.5 ! (n) $) ;
  \draw[-<,>=latex,line width=0.01pt] ($ (b) ! 0.4 ! (n) $) -- ($ (b) ! 0.5 ! (n) $) ;
  \draw[->,>=latex,line width=0.01pt] ($ (c) ! 0.4 ! (n) $) -- ($ (c) ! 0.5 ! (n) $) ;
  \draw[-<,>=latex,line width=0.01pt] ($ (d) ! 0.4 ! (n) $) -- ($ (d) ! 0.5 ! (n) $) ;}
\fi
\end{tikzpicture}
}

\newcommand{\threeTet}[6]{
\begin{tikzpicture}[y=0.80pt, x=0.8pt,yscale=-1, inner sep=0pt, outer sep=0pt,scale=#6]
  \path[draw=black,line join=round,line cap=round]
    (300.0000,972.3622) -- (190.0000,1002.3622) -- (140.0000,872.3622) --
    (250.0000,772.3622) -- (340.0000,862.3622) -- cycle;
  \path[draw=black,line join=round,line cap=round]
    (250.0000,772.3622) -- (300.0000,972.3622) -- (140.0000,872.3622);
  \path[draw=black,line join=round,line cap=round]
    (140.0000,872.3622) -- (219.2839,868.3980)(230.3456,867.8449) --
    (267.4849,865.9879)(279.6331,865.3805) -- (340.0000,862.3622) --
    (289.6211,909.3825)(280.4793,917.9149) --
    (257.2290,939.6151)(248.7351,947.5428) -- (190.0000,1002.3622) --
    (210.7952,922.6473)(213.2870,913.0954) --
    (218.0316,894.9077)(220.1100,886.9406) -- (250.0000,772.3622);
  \path[fill=black] (307,978) node[above right] {\scalebox{0.65}{$#1$}};
  \path[fill=black] (335,852.36218) node[above right]  {\scalebox{0.65}{$#2$}};
  \path[fill=black] (265,782.36218) node[above right]  {\scalebox{0.65}{$#3$}};
  \path[fill=black] (125,867.36218) node[above right]  {\scalebox{0.65}{$#4$}};
  \path[fill=black] (156,1002.3622) node[above right] {\scalebox{0.65}{$#5$}};
\end{tikzpicture}}

\newcommand{\twoTet}[6]{
\begin{tikzpicture}[y=0.80pt, x=0.8pt,yscale=-1, inner sep=0pt, outer sep=0pt,scale=#6]
  \path[draw=black,line join=round,line cap=round]
    (300.0000,972.3622) -- (190.0000,1002.3622) -- (140.0000,872.3622) --
    (250.0000,772.3622) -- (340.0000,862.3622) -- cycle;
  \path[draw=black,line join=round,line cap=round]
    (250.0000,772.3622) -- (300.0000,972.3622) -- (140.0000,872.3622);
  \path[draw=black,line join=round,line cap=round]
    (279.6331,865.3805) -- (340.0000,862.3622) -- (289.6211,909.3825);
  \path[draw=black,line join=round,line cap=round]
    (280.4793,917.9149) -- (257.2290,939.6151);
  \path[draw=black,line join=round,line cap=round]
    (248.7351,947.5428) -- (190.0000,1002.3622);
  \path[draw=black,line join=miter,line cap=round]
    (219.2839,868.3980) -- (230.3456,867.8449)(230.3456,867.8449) --
    (267.4849,865.9879)(140.0000,872.3622) -- (219.2839,868.3980);
  \path[fill=black] (307,978) node[above right] {\scalebox{0.65}{$#1$}};
  \path[fill=black] (335,852.36218) node[above right]  {\scalebox{0.65}{$#2$}};
  \path[fill=black] (265,782.36218) node[above right]  {\scalebox{0.65}{$#3$}};
  \path[fill=black] (125,867.36218) node[above right]  {\scalebox{0.65}{$#4$}};
  \path[fill=black] (156,1002.3622) node[above right] {\scalebox{0.65}{$#5$}};

\end{tikzpicture}}

\newcommand{\torusCube}[9]{
\begin{tikzpicture}[y=0.80pt, x=0.8pt,yscale=-1, inner sep=0pt, outer sep=0pt,scale=#9]
  \begin{scope}[cm={{0.08284,0.0,0.0,0.08284,(38.62164,986.2694)}}]
    \begin{scope}[cm={{1.06299,0.0,0.0,-1.06299,(-186.02362,789.27165)}},draw=black,line join=miter,line cap=butt,miter limit=10.43]
    \end{scope}
  \end{scope}
  \path[draw=black,line join=round,line cap=round,miter limit=4.00,line
    width=0.400pt] (50.2500,1032.1122) -- (50.2500,1002.1122);
  \path[draw=black,line join=round,line cap=round,miter limit=4.00,line
    width=0.400pt] (20.2500,1032.1122) -- (20.2500,1002.1122);
  \path[draw=black,line join=round,line cap=round,miter limit=4.00,line
    width=0.400pt] (50.2500,1002.1122) -- (20.2500,1002.1122);
  \path[draw=black,line join=round,line cap=round,miter limit=4.00,line
    width=0.400pt] (30.2500,992.1122) -- (20.2500,1002.1122);
  \path[draw=black,line join=round,line cap=round,miter limit=4.00,line
    width=0.400pt] (60.2500,992.1122) -- (50.2500,1002.1122);
  \path[draw=black,line join=round,line cap=round,miter limit=4.00,line
    width=0.400pt] (30.2500,992.1122) -- (60.2500,992.1122);
  \path[draw=black,line join=round,line cap=round,miter limit=4.00,line
    width=0.400pt] (60.2500,1022.1122) -- (60.2500,992.1122);
  \path[draw=black,line join=round,line cap=round,miter limit=4.00,line
    width=0.400pt] (50.2500,1032.1122) -- (60.2500,1022.1122);
  \path[draw=black,line join=round,line cap=round,miter limit=4.00,line
    width=0.200pt] (30.2500,992.1122) -- (30.2500,1001.0747)(30.2500,1003.3476) --
    (30.2500,1010.4186)(30.2500,1013.9542) -- (30.2500,1022.1122);
  \path[draw=black,line join=round,line cap=round,miter limit=4.00,line
    width=0.200pt] (60.2500,1022.1122) -- (51.5400,1022.1122)(49.0146,1022.1122)
    -- (41.4385,1022.1122)(38.6605,1022.1122) -- (30.2500,1022.1122);
  \path[draw=black,line join=round,line cap=round,miter limit=4.00,line
    width=0.200pt] (20.2500,1032.1122) -- (30.2500,1022.1122);
  \path[draw=black,line join=round,line cap=round,miter limit=4.00,line
    width=0.400pt] (20.2500,1002.1122) -- (50.2500,1032.1122);
  \path[draw=black,line join=round,line cap=round,miter limit=4.00,line
    width=0.400pt] (50.2500,1002.1122) -- (60.2500,1022.1122);
  \path[draw=black,line join=round,line cap=round,miter limit=4.00,line
    width=0.400pt] (30.2500,992.1122) -- (50.2500,1002.1122);
  \path[draw=black,line join=round,line cap=round,miter limit=4.00,line
    width=0.200pt] (20.2500,1002.1122) -- (30.2500,1022.1122);
  \path[draw=black,line join=round,line cap=round,miter limit=4.00,line
    width=0.200pt] (30.2500,1022.1122) -- (50.2500,1032.1122);
  \path[draw=black,line join=round,line cap=round,miter limit=4.00,line
    width=0.200pt] (30.2500,992.1122) -- (39.4416,1001.3038)(41.2094,1003.0716) --
    (44.4292,1006.2915)(46.1970,1008.0592) --
    (49.2275,1011.0897)(51.3109,1013.1731) -- (60.2500,1022.1122);
  \path[draw=black,line join=round,line cap=round,miter limit=4.00,line
    width=0.200pt] (30.2500,1022.1122) -- (34.4305,1017.9317)(36.0089,1016.3533)
    -- (50.2500,1002.1122);
  \path[fill=black] (17.399113,1032.236) node[above right]  {\scalebox{0.65}{$#1$}};
  \path[fill=black] (53.447334,1032.1368) node[above right] {\scalebox{0.65}{$#2$}};
  \path[fill=black] (25.25,1022.1121) node[above right] {\scalebox{0.65}{$#3$}};
  \path[fill=black] (61.492386,1022.1121) node[above right]  {\scalebox{0.65}{$#4$}};
  \path[fill=black] (17.328045,1003.2214) node[above right]  {\scalebox{0.65}{$#5$}};
  \path[fill=black] (52.366116,1002.9758) node[above right] {\scalebox{0.65}{$#6$}};
  \path[fill=black] (24.891497,994.84943) node[above right]{\scalebox{0.65}{$#7$}};
  \path[fill=black] (61.156353,994.97632) node[above right] {\scalebox{0.65}{$#8$}};
  \path[draw=black,line join=round,line cap=round,miter limit=4.00,line
    width=0.400pt] (20.2500,1032.1122) -- (50.2500,1032.1122);
  \path[fill=black] (57.4,1010) node[above right] {\scalebox{0.65}{$k$}};
  \path[fill=black] (44,1021) node[above right] {\scalebox{0.65}{$g$}};
  \path[fill=black] (57,1029.0608) node[above right] {\scalebox{0.65}{$h$}};
  \path[fill=black] (44,990) node[above right] {\scalebox{0.65}{$g^{'}$}};
   \path[fill=black] (20,998.0608) node[above right] {\scalebox{0.65}{$h'$}};
 \path [gray,fill, opacity=0.4, opacity=0.3] (20.2500,1032.1122) -- (50.2500,1032.1122) -- (60.2500,1022.1122) -- (30.2500,1022.1122)--cycle;
\path [lightgray,fill, opacity=0.4, opacity=0.3] (20.2500,1002.1122) -- (50.2500,1002.1122) -- (60.2500,992.1122) -- (30.2500,992.1122)--cycle;
\node[below] at (37,1035.0608) {$K_1$};
\node[above] at (37,990.0608) {$K_2$};
\end{tikzpicture}
}

\newcommand{\tII}{\text{II}}
\newcommand{\tIII}{\text{III}}

\newcommand{\cH}{H}

\title{Gapped Boundary Theory of the Twisted Gauge Theory Model of Three-Dimensional Topological Orders}
\date{\today}

\author[a,b]{Hongyu Wang}
\author[a,b]{Yingcheng Li}
\author[c]{Yuting Hu}
\author[a,b,c,d,e,1]{Yidun Wan,\note{Corresponding author}}
\affiliation[a]{State Key Laboratory of Surface Physics, Fudan University, Shanghai 200433, China}
\affiliation[b]{Department of Physics and Center for Field Theory and Particle Physics, Fudan University, Shanghai 200433, China}
\affiliation[c]{Department of Physics and Institute for Quantum Science and Engineering, Southern University of Science and Technology, Shenzhen 518055, China}
\affiliation[d]{Institute for Nanoelectronic devices and Quantum computing, Fudan University, Shanghai 200433, China}
\affiliation[e]{Collaborative Innovation Center of Advanced Microstructures, Nanjing, 210093, China}

\emailAdd{ydwan@fudan.edu.cn}

\abstract{
We extend the twisted gauge theory model of topological orders in three spatial dimensions to the case where the three spaces have two dimensional boundaries. We achieve this by systematically constructing the boundary Hamiltonians that are compatible with the bulk Hamoltonian. Given the bulk Hamiltonian defined by a gauge group $G$ and a four-cocycle $\omega$ in the fourth cohomology group of $G$ over $U(1)$, a boundary Hamiltonian can be defined by a subgroup $K$ of $G$ and a three-cochain $\alpha$ in the third cochain group of $K$ over $U(1)$. The boundary Hamiltonian to be constructed must be gapped and invariant under the topological renormalization group flow (via Pachner moves),  leading to a generalized Frobenius condition.  Given $K$, a solution to the generalized Frobenius condition specifies a gapped boundary condition. We derive a closed-form formula of the ground state degeneracy of the model on a three-cylinder, which can be naturally generalized to three-spaces with more boundaries. We also derive the explicit ground-state wavefunction of the model on a three-ball. The ground state degeneracy and ground-state wavefunction are both presented solely in terms of the input data of the model, namely, $\{G,\omega,K,\alpha\}$.
}

\begin{document}

\maketitle
\flushbottom

\section{Introduction}\label{sec:intro}
Systems bearing intrinsic topological orders in two\cite{Wen1989,Wen1989a,Wen1990a,Wen1990c,Kitaev2003a,Levin2004,Kitaev2006,Chen2012a,Levin2012,Hung2012,Hu2012,Hu2012a,Mesaros2011,Lin2014,Kong2014,Lan2018a} and three\cite{Walker2011,Wang2014b,Jiang2014a,Wang2014c,Kong2014,Wan2014,Wang2015,Bullivant2016,Bullivant2017a,Lan2017,Delcamp2017,Delcamp2018a,Lan2018,Cheng2018} spatial dimensions have received substantial attention recently because they not only have greatly expanded and deepened our understanding of phases of matter but also  have practical applications, in particular certain 2-dimensional topologically ordered systems can support topological quantum computation\cite{Kitaev2003a,Nayak2008}. Celebrated candidates of two-dimensional topological phases include chiral spin liquids\cite{Kalmeyer1987,Wen1989a}, $\Z_2$ spin liquids\cite{Read1991,Wen1991,Moessner2001}, Abelian quantum Hall states\cite{Klitzing1980,Tsui1982,Laughlin1983}, and non-Abelian fractional quantum Hall states\cite{TaoWu1984,Moore1991,Wen1991b,Willett1987,Radu2008}.

Although three is our physical spatial dimension, our understanding of three-dimensional topological orders is far behind that of the two-dimensional ones. Moreover, candidates of three-dimensional topological orders are still missing. Recent works also shows that three-dimensional topological orders are related to quantum gravity\cite{Dittrich2017,Delcamp2018}. It is therefore demanding and interesting to study theories and models of three-dimensional topological orders. Symmetry considerations have led to the twisted gauge theory (TGT) model that describes and classifies a large class of three-dimensional topological orders\cite{Jiang2014a,Wan2014}. The TGT model is an exactly solvable lattice Hamiltonian extension of the $(3+1)$-dimensional Dijkgraaf-Witten topological gauge theory with general finite gauge groups on a lattice. The TGT model is specified by a Hamiltonian $H^{G,\omega}$ defined on a triangulation of a $3$-manifold, where $G$ is a finite gauge group, and the usual gauge transformation is twisted by a $U(1)$ \fc $\omega\in H^4[G,U(1)]$, the fourth cohomology group of $G$. This model however cannot handle the situation where the $3$-manifolds have boundaries, whereas realistic materials---in particular in three dimensions---mostly do have boundaries. 
 
When boundaries are present, the Hamiltonian of the model should contain boundary terms too. For the model to be able to describe topological orders with gapped boundaries\footnote{So as the topological orders to have well defined ground state degeneracy.}, we only consider the boundary terms that can render the boundaries gapped. Such boundary terms may affect the spectrum of the model in two aspects. First, the topologically protected ground state degeneracy (GSD) of a topological order may be modified due to its gapped boundary conditions. Second, different gapped boundary conditions correspond to different sets of bulk excitations condensing at the boundaries\cite{Levin2013,Kong2013,HungWan2013a,HungWan2014,Kong2017}. These two aspects had been an open problem of topological orders in two and three dimensions. Until only recently these two problems in two-dimensions have been solved for Abelian topological orders\cite{Wang2012} and for general, non-Abelian topological orders\cite{HungWan2014,Lan2014,Bullivant2017} but are still open for three-dimensions. 

More broadly, on the one hand, a dynamical theory incapable of handling boundary conditions is not a complete dynamical theory. Therefore, the TGT model is yet a complete dynamical model on open three-manifolds. On the other hand, the bulk and boundary relation in light of anyon condensation (in any dimensions\cite{Kong2017}) offers a new perspective of of holographic principle, which may shed new light on the AdS/CFT correspondence.\footnote{Via private communication with Ling-Yan Hung.}     

Recently, there has been a couple works\cite{Wang2016,Kong2017,Wang2017,Fidkowski2017} dealing with the boundary theories of topological orders in two and higher dimensions. They either take an algebraic approach\cite{Wang2016,Kong2017}, or studies a particular interesting example\cite{Fidkowski2017}, or focusing on symmetry (protected) enriched topological orders\cite{Wang2017}. Nevertheless, we shall generalize the systematic approach of exactly solvable Hamiltonian developed by two of us in Refs. \cite{Bullivant2017,Hu2017,Hu2017a}, where the twisted quantum double model and the string-net model of two-dimensional topological orders are extended to include boundaries. In this paper, we extend the three-dimensional TGT model to include boundaries. We shall restrict the boundary degrees of freedom in the extended TGT model to take value in $K\subseteq G$. We then add to the original TGT Hamiltonian compatible boundary terms that depend on $K$ and a $3$-cochain $\alpha\in C^3[K,U(1)]$. Studying this extended TGT model leads to the following main results.   

 We extend the TGT bulk Hamiltonian by a local boundary Hamiltonian with boundary degrees of freedom being in a subgroup (not necessarily a proper one) of the gauge group in the bulk. The local operators in boundary Hamiltonian are constructed in terms of a $3$-cochain of the boundary subgroup. 
The boundary local Hamiltonian needs to be exactly solvable to make sure that the model describes gapped boundaries. The boundary Hamiltonian should also be compatible with the bulk Hamiltonian, such that the ground states are invariant under topology-preserving mutation of triangulation both in the bulk and on the boundary. These requirements will result in what we call the generalized Frobenius condition (as a generalization of the two-dimensional version\cite{Bullivant2017}) that dictates the gapped boundary conditions. 

Our extended Hamiltonian has the advantage that enables us to write down an explicit ground-state wavefunction of our model on a $3$-ball and a closed-form formula for the GSD on a $3$-cylinder, both are cast in terms of the input data of the model only, namely, $\{G,\omega,K,\alpha\}$. We show a couple of examples.

The paper is organized as follows. Section \ref{sec:modelRev} briefly reviews the TGT model. Sections \ref{sec:bdryHamiltonian} and \ref{sec:one2one} systematically constructs the boundary Hamiltonian. Section \ref{sec:groundstate} derives the explicit ground-state wavefunction of our model on a $3$-ball. Section \ref{sec:GSD} deduces the closed-form formula of the ground state degeneracy (GSD) of our model on a $3$-cylinder. The Appendices include a basic introduction to group cohomology and certain details too much to appear in the main text.
\section{Review of the TGT mode}\label{sec:modelRev}
\begin{figure}[!h]
\centering
  \includegraphics[scale=0.6]{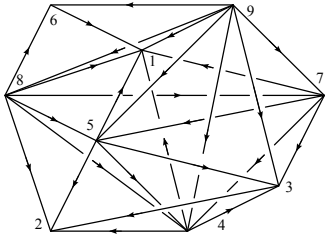}
  \caption{A portion of a graph that represent the basis vectors in the Hilbert space. Each edge carries an arrow and is assigned a group element denoted by $[ab]$ with $a<b$.}
  \label{fig:GraphConfiguration}
\end{figure}

  Here we briefly review the TGT model of topological orders on closed 3-manifold. The TGT model is defined by a low-energy effective Hamiltonian $H_{G,\omega}$ on a triangulation $\Gamma$ (see Fig.\ref{fig:GraphConfiguration}) of  a closed, orientable, $3$-manifold, e.g., a $3$-sphere and a $3$-torus. Each edge $ab$ from vertex $a$ to vertex $b$ in $\Gamma$ is graced with a group element $[ab]\in G$, rendering the Hilbert space of the model consists of all possible configurations of the group elements on the edges of $\Gamma$. Namely,
\be
\Hil_{\Gamma,G}=\{[i j]\in G|i,j\in V(\Gamma)\},
\ee
where $V(\Gamma)$ is the set of vertices of $\Gamma$. 
The states are orthogonal obviously. The group elements on the edges can be considered as the discretized gauge field of the underlying Dijkgraaf-Witten topological gauge theory. The graph is oriented with an arbitrary ordering of the vertices\footnote{The corresponding triangulation is said to have a branching structure.}, such that each edge is arrowed from its larger vertex to the smaller and  that $[ab]=\overline{[ba]}$, where the overline denotes the group inverse. A vertex ordering is called an enumeration\cite{Hu2012} and does not affect the physics as long as the relative order of the vertices persists when the graph mutates. A graph $\Gamma$ mutates via the $3$-dimensional Pachner moves\cite{Pachner1978,Pachner1987}, seen in Eq. \eqref{eq:Pachner}.  
\be\label{eq:Pachner}
\begin{aligned}
& f_1:\; \bmm\twoTet{}{}{}{}{}{0.25}\emm\mapsto\bmm\threeTet{}{}{}{}{}{0.25}\emm,\;\\
& f_2:\; \bmm\threeTet{}{}{}{}{}{0.25}\emm\mapsto\bmm\twoTet{}{}{}{}{}{0.25}\emm,\;\\
& f_3:\; \bmm\fourTet{}{}{}{}{0}{}{0.5}{0}\emm\mapsto\bmm\fourTet{}{}{}{}{1}{}{0.5}{0}\emm,\\
& f_4:\; \bmm\fourTet{}{}{}{}{1}{}{0.5}{0}\emm\mapsto\bmm\fourTet{}{}{}{}{0}{}{0.5}{0}\emm.
\end{aligned}
\ee

\begin{figure}[h!]
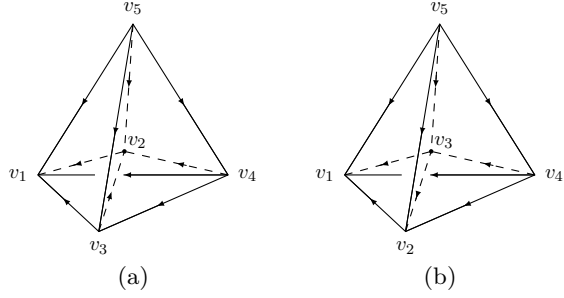

\centering
\subfigure[]{\fourTet{v_1}{v_3}{v_5}{v_4}{2}{v_2}{1}{1} \label{fig:4cocycleA}}
\subfigure[]{\fourTet{v_1}{v_2}{v_5}{v_4}{3}{v_3}{1}{1} \label{fig:4cocycleB}}
\caption{(a) The defining graph of the \fc $[v_1v_2 v_3v_4v_5]$. (b) Of $[v_1v_2v_3v_4v_5]^{-1}$.}
\label{fig:4cocycle}
\end{figure}

The mutation of $\Gamma$ turns $\Gamma$ into a different graph $\Gamma'$ and hence alters the total Hilbert space of the model. Nevertheless, as shown in Ref.\cite{Hu2012}, the topological properties of the topological order described by the model $H_{G,\omega}$ do not change because mutations cannot alter the topology of the surface.
For simplicity, we neglect drawing the group elements on the edges but keep only the vertex labels. We may also  refer to $ abc$ as a $2$-simplex, $abcd$ as a $3$-simplex, and $[abcd]$ as a $3$-cocycle on the tetrahedron . On any part of $\Gamma$ that resembles Fig. \ref{fig:4cocycleA}, one can define a  normalized $4$-cocycle $[v_1v_2v_3v_4v_5]\in H^4[G,U(1)]$. The four group variables in the $ 4$-cocycle from left to right are respectively $[v_1v_2]$, $[v_2v_3]$ , $[v_3v_4]$, and $[v_4v_5]$, which are along the path from the least vertex $v_1$ to the greatest vertex $v_5$ passing $v_2$ and $v_3$ in order. Appendix A reviews necessary rudiments of group cohomology. Here one should keep in mind that a 4-cocycle is an equivalence class of $U(1)$-valued functions on $G^4=G\x G\x G\x G$. A normalized $4$-cocycle is a particular representative that satisfies the normalization condition 
\begin{align}
  \label{NormalizationCondition}
  [v_1v_2v_3v_4v_5]=1    \text{  if $[v_iv_{i+1}]=1\in G$ for any $i$}
\end{align}
and the $4$-cocyle condition
\begin{align}
  \label{4CocycleCondition}
  \Pi^{5}_{i=1}(-1)^{i+1}[v_{1}...\hat v_{i}...v_{5}]=1,
\end{align}
where $\hat v_{i}$ means removing the point $v_i$. We take this notation in order to figure out easily the triangles required to be flat  (defined in Eq. \eqref{eq:chainRule}) when we use Pachner moves. To make sure all the elements multiplication is allowed, one need to check if all the triangles $v_{i-1}v_{i}v_{i+1}$($i=2,\dots,4$) are flat Eq. \eqref{eq:chainRule}. 

Shown in Ref.\cite{Hu2012}, each $4$-cocycle defines a $3$-dimensional topological order and the choice of the normalized $4$-cocycle as the representative is merely a convenience that does not affect the physics. A graph like Fig. \ref{fig:4cocycleA} has an  associated $4$-cocyle whose orientation  determined as follows. One first reads off a list of the  vertices  from any of the four tetrahedra of the defining graph of the $4$--cocycle, e.g., the $v_{2}v_{3}v_{4}v_{5}$ from Fig. \ref{fig:4cocycleA} and $v_{3}v_{2}v_{4}v_{5}$ from Fig. \ref{fig:4cocycleB}. One then appends the remaining vertex to the beginning of the list,  e.g., the $v_{1}v_{2}v_{3}v_{4}v_{5}$ from Fig. \ref{fig:4cocycleA} and $v_{1}v_{3}v_{2}v_{4}v_{5}$ from Fig. \ref{fig:4cocycleB}. If the list can be reordered ascendingly by even permutations, such as $v_{1}v_{2}v_{3}v_{4}v_{5}$ from Fig. \ref{fig:4cocycleA}, one has the $4$-cocycle $[v_{1}v_{2}v_{3}v_{4}v_{5}]$, otherwise $[v_{1}v_{2}v_{3}v_{4}v_{5}]^{-1}$  as by the $v_{1}v_{3}v_{2}v_{4}v_{5}$ from Fig. \ref{fig:4cocycleB}.

 The Hamiltonian of our model reads
\be\label{eq:Hamiltonian}
 H_{G,\omega}=-\sum_v A_v-\sum_f B_f,
\ee
where $B_f$ are the face operators defined on the triangular faces $f$, and $A_v$ are the vertex operators defined at the vertices $v$. The operator $B_f$ acts on the face $f$ in a basis state vector as
\begin{align}
  \label{eq:actionOfBf}
  B_f\BLvert \oneTriangle{v_1}{v_2}{v_3} \Brangle
  =\delta_{[v_1v_2]\cdot[v_2v_3]\cdot[v_3v_1]}
  \BLvert \oneTriangle{v_1}{v_2}{v_3}\Brangle .
\end{align}
The discrete
delta function $\delta_{[v_1v_2]\cdot[v_2v_3]\cdot[v_3v_1]}$ is unity if ${[v_1v_2]\cdot[v_2v_3]\cdot[v_3v_1]=1 }$, where $1$ is the identity element in $G$, and 0 otherwise. Note again that here, the ordering of $v_1,v_2$, and $v_3$ is irrelevant because of the identities
$\delta_{[v_1v_2]\cdot[v_2v_3]\cdot[v_3v_1]}
=\delta_{[v_3v_1]\cdot[v_1v_2]\cdot[v_2v_3]}$ and
$\delta_{[v_1v_2]\cdot[v_2v_3]\cdot[v_3v_1]}
=\delta_{\{\overline{[v_1v_2]\cdot[v_2v_3]\cdot[v_3v_1]\}}}
=\delta_{\overline{[v_3v_1]}\cdot\overline{[v_2v_3]}\cdot\overline{[v_1v_2]}}
=\delta_{[v_1v_3]\cdot[v_3v_2]\cdot[v_2v_1]}$.
In other words,
if $B_f=1$ on a face $f$, the three group elements on the three edges of $f$ are related by a \textit{chain
rule}:
\begin{equation}
\label{eq:chainRule}
[v_1v_3]=[v_1v_2]\cdot[v_2v_3]
\end{equation}
for any enumeration $v_1,v_2,v_3$ of the three vertices of the face $f$. The chain rule \eqref{eq:chainRule} is physically known as the \textit{flatness condition} in the sense that the gauge connection along the edges of a triangular face is flat. The operator $A_v$ acting on a vertex $v$ is an average
\begin{equation}
  \label{eq:Av}
  A_v=\frac{1}{|G|}\sum_{[vv']=g\in G}A_v^g
\end{equation}
over the operators $A_v^g$ specified by a group element $g\in G$ acting on the same vertex. The action of $A^g_v$ replaces $v$ by a new enumeration $v'$ that is less than $v$ but greater than all the vertices that are less than $v$ in the original set of enumerations before the action, such that $[v'v]=g$. In a dynamical language, $v'$ may be thought as on the next \textquotedblleft time" slice, and there is an edge $[v'v]\in G$ in the $(3+1)$-dimensional \textquotedblleft spacetime" picture. We illustrate such an action in the example below, which suffices to show the general definition on arbitrary vertex.
\be
A_{v_4}^g : \BLvert \fourTet{v_1}{v_2}{v_4}{v_3}{0}{}{0.5}{1} \Brangle \mapsto \BLvert\fourTet{v_1}{v_2}{v'_4}{v_3}{0}{}{0.5}{1}\Brangle,
\ee
where on the RHS, the new enumerations are in the order $v_1<v_2< v_3<v'_4<v_4$, satisfying the following flatness conditions.
\be
\begin{aligned}
\label{eq:ChainRuleInBph}
&[v_1{v'_4}]=[v_1v_4]\cdot[v_4{v'_4}],\\
&[v_2{v'_4}]=[v_2v_4]\cdot[v_4{v'_4}],\\
&[v_3{v'_4}]=[v_3v_4]\cdot[v_4v'_4].
\end{aligned}
\ee

Now imagine to put together the two tetrahedra before and after the action of $A^g_{v_4}$ as two spatial slices connected by the `temporal' edge $v'_4v_4$, which is not shown, we obtain a $4$-simplex. This is a picture of time evolution, which motivates us to attribute the amplitude of $A^g_{v_4}$ to an evaluation of the $4$-simplex. This amplitude is naturally given by the \fc associated with the $4$-simplex (recall our earlier discussion). That is,

\be\label{eq:AvAmpSingleTet}
\begin{aligned}
&\Blangle\fourTet{v_1}{v_3}{v'_4}{v_3}{0}{}{0.5}{1}\BRvert A^g_{v_4}\BLvert  \fourTet{v_1}{v_2}{v_4}{v_3}{0}{}{0.5}{1} \Brangle\\
 =\ & \delta_{[v'_4v_4],g}\Blangle\fourTet{v_1}{v_2}{v_4}{v_3}{4}{v'_4}{0.6}{1} \Brangle\\
=\ &\delta_{[v'_4v_4],g}[v_1v_2v_3v'_4v_4]^{\epsilon(v_1v_2v_3v'_4v_4)}\\
=\ &\delta_{[v'_4v_4],g}[v_1v_2v_3v'_4v_4]^{-1}
\end{aligned}
\ee

\begin{figure}[h!]
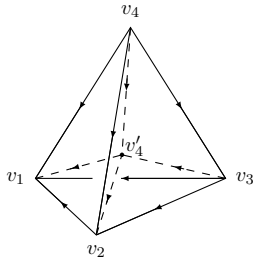

\centering
\BvTri{v_1}{v_2}{v_4}{v_3}{3}
\caption{The topology of the action of $A_{v_4}^g$.}
\label{fig:BvTrivalent}
\end{figure}

The $4$-cocycles appearing on the RHS of Eq. (\ref{eq:AvAmpSingleTet}) are understood from Fig. \ref{fig:BvTrivalent}.  This figure illustrates the time evolution of the graph due to the action of $A^g_{v_4}$. To define $\epsilon(v_1v_2v_3v'_4v_4)$, we first need to define the orientation of tetrahedron, $\epsilon(v_1v_2v_{3}v_4)$. The convention is as follows.\begin{convention}\label{conv:TetOrientation}
One can grab the triangle of $v_1v_2v_{3}v_4$ that does not contain the largest vertex, i.e., the triangle $v_1v_2v_{3}$, along its boundary, such that the three vertices are in ascending order, while the thumb points to the vertex $v_{4}$. If one must use one's right hand to achieve this, $v_1v_2v_{3}v_4$'s orientation is $+$, or $\epsilon(v_1v_2v_{3}v_4)=1$, otherwise $-$, or $\epsilon(v_1v_2v_{3}v_4)=-1$.    
\end{convention}
We then have the following definition of $\epsilon(v_1v_2v_3v'_4v_4)$, \begin{convention}\label{conv:4cocyleAv1Tet}
Since the new vertex $v'_4$ is set to be slightly off the 3d space of the tetrahedron $v_1v_2v_3v_4$, and since every newly created vertex bears a label slightly less than that of the original vertex acted on by the vertex operator, one can always choose the convention such that $\epsilon(v_1v_2v_3v'_4v_4) =\epsilon(v_1v_2v_3v_4)\sgn{v'_4,v_1,v_2,v_3,v_4}$. And $\sgn{v'_4,v_1,v_2,v_3,v_4}$ is the sign of the permutation that takes the list of vertices in the argument to purely ascending as $(v_1,v_2,v_3,v'_4,v_4,v_5)$, which embraces the 4-simplex $v_1v_2v_3v'_4v_4$. 
\end{convention} 

The vertex operator in Eq. \eqref{eq:AvAmpSingleTet} can naturally extends its definition from a trivalent vertex to a vertex $v$ of any valence. The number of $4$--cocyles in the phase factor brought by the action of $A_v^g$ on $v$ equals $v$'s valence of. It is clear that $A^{g=1}_v\equiv \mathbb{I}$ by the normalization of $4$-cocycles. The Hamiltonian \eqref{eq:Hamiltonian} is exactly solvable when all the triangles are flat, namely, in the Hilbert subspace $\Hil^{B_f=1}$. The ground states are the common $+1$ eigenvectors of these projectors. On a $3$-torus,  a ground-state basis states  can be labeled by $\ket{A,B,\mu}$, where $A$ is a conjugacy class of $G$, $B$ is a conjugacy class in the centralizer subgroup $Z^A$, and $\mu$ an irreducible representation of the centralizer of  both $A$ in $G $ and $B$ in $Z^A$. When both $A$ and $B$ are nontrivial, the excitations are loops instead of points because $A$ and $B$ are the bases of fluxes in two directions. The ground states $\ket{A,B,\mu}$ would cease to form a quantum-double algebra. The representations $\mu$ are of a special type, called $\beta_{k^A,g^B}$-regular. These $\beta_{k^A,g^B}$ are doubly-twisted $2$-cocycles derived from $\omega$ via the slant product (see Appendix 1). Interestingly, the topological orders described by the TGT model are not classified by the $4$-cocycles $w\in H^4[G,U(1)]$ given $G$ but instead classified by the doubly-twisted $2$-cocycles $\beta_{k^A,g^B}$ derived from $\omega$.\cite{Hu2012} On a 3-torus, the GSD of the model $H_{G,\omega}$ is
\be\label{eq:GSDrepresentations}
  \text{GSD}
  =\sum_{A}\#(\beta_{k^A,g^B}-\text{regular irreps of }Z^{A,B} ),
\ee     
where the sum runs over all the conjugacy classes of $G$ and $Z^{A,B}\subset G$ is the centralizer of both $A$ and $B$.

Unlike $2+1$-dimensional TQD models on a torus, the ground states and quasi-excitations don't remain in one-to-one correspondence for $3+1$-dimensional TGT models on a $3$-torus.

\section{The TGT model with boundaries}\label{sec:bdryHamiltonian}
\begin{figure}[!h]
        \centering
        \includegraphics[scale=0.15]{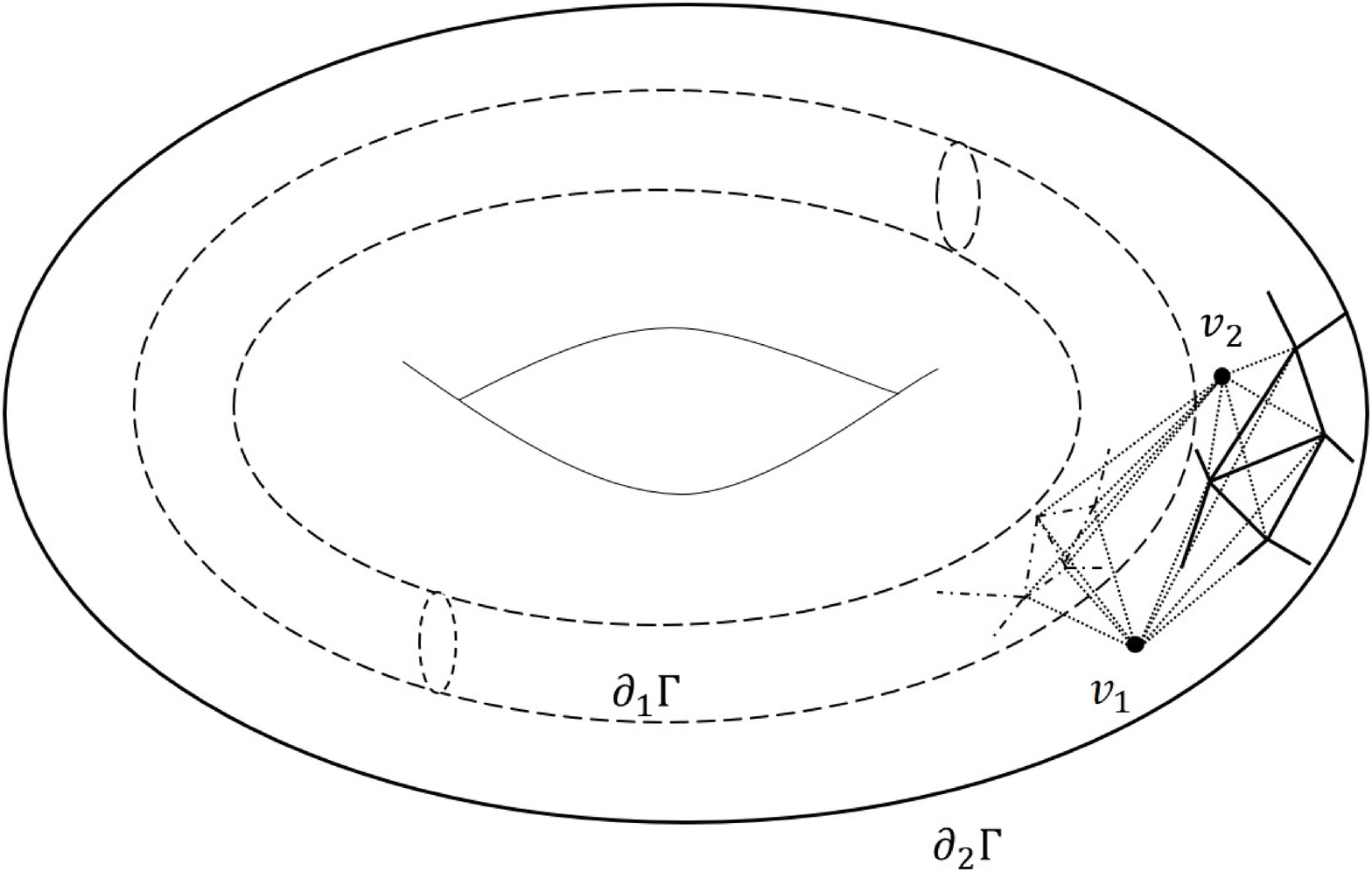}
        \caption{A $3$-cylinder, i.e., a solid torus with a solid torus removed from interior. This is an example of a $3$-manifold with two boundaries, the inner boundary (dashed line) $\partial_1\Gamma$ and the outter boundary (solid line) $\partial_2\Gamma$. A portion of the triangulation $\Gamma$ of this $3$-cylinder is shown, where $v_1$ and $v_2$ are two vertices in the bulk.} 
        \label{fig:multiboundary}
\end{figure}
We now try to extend the TGT model reviewed above to one that is defined on a $3$-dimensional space with boundaries (see Fig. \ref{fig:multiboundary} for an example). We do not distinguish multiple boundaries from a single boundary that has multiple components. 

\begin{figure}[!h]
        \centering
        \includegraphics[scale=0.15]{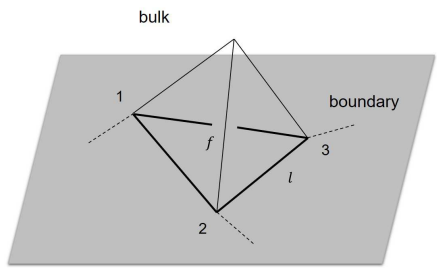}
        \caption{The boundary is the grey plane, above which is the bulk. A bulk tetrahedron and  a boundary face $f=123$ is explicitly shown. The dashed lines represent the rest of the graph that is now drawn.}
        \label{fig:boundaryface}
\end{figure}

Let us consider a portion of boundary graph shown in Fig. \ref{fig:boundaryface} with three boundary vertices $1,2,3$. There is a boundary plaquette $f=123$ with all the three edges on the boundary. 

We first need to specify the degrees of freedom living on the boundaries. Since we require the boundaries to be gapped for the topological GSD to be well defined, the boundary degrees of freedom would naturally arise from those in the bulk. To be precise, consider a single boundary $\partial\Gamma$ of a graph $\Gamma$, if the bulk degrees of freedom take value in a finite group $G$, the degrees of freedom on $\partial\Gamma$ take value in a subgroup $K\subseteq G$. As in the $(2+1)$-dimensional cases, $K$ is allowed to be the trivial subgroup $\{1\}$ and $G$ itself too. If $\Gamma$ has $M$ boundaries $\partial_1\Gamma,\ \partial_2\Gamma,\dots,\partial_M\Gamma$, the subgroups $K_1,\ K_2,\dots,K_n\subseteq G$ are not necessarily the same or different. 
Having set up the boundary degrees of freedom, we can now propose the Hamiltonian of the TGT model with multiple boundaries:
\be\label{eq:HamWithBoundaries}       
H^{K,\alpha}_{G,\omega}=H_{G,\omega}-\sum_{i=1}^M\left(\sum_{v\in\partial_i\Gamma}
A_v^{K_i}-\sum_{f\in\partial_i\Gamma}B_f^{K_i}-\sum_{l\in\partial_i\Gamma}C_l^{K_i}\right),
\ee
where $H_{G,\omega}$ is the bulk Hamiltonian \eqref{eq:Hamiltonian}, and the rest are the boundary terms to be explained in order.

An operator $B_f^K$ acts on a boundary plaquette $f$ and is defined as
in the example below.\be\label{Bf1}
\begin{aligned}
B_{123}^K\BLvert\fourTetboundary{1}{2}{}{3}{0}{}{0.5}{0}\Brangle=\delta_{[12]\cdot[23]\cdot[31]}\BLvert\fourTetboundary{1}{2}{}{3}{0}{}{0.5}{0}\Brangle.
\end{aligned}
\ee
This $\delta$ function is $1$ if $[12]\cdot[23]\cdot[31]=1\in K$ and $0$ otherwise. An operator $C_l^K$ projects the degrees of freedom on the boundary edge $l$ into the subgroup $K$, namely,
\be\label{Bf2}
\begin{aligned}
&C_{l}^K=\sum_{k\in K}C_{l}^k,\\
&C_{23}^k\BLvert\fourTetboundary{1}{2}{}{3}{0}{}{0.5}{0}\Brangle=\delta_{[23],k}\BLvert\fourTetboundary{1}{2}{}{3}{0}{}{0.5}{0}\Brangle.
\end{aligned}
\ee

Clearly, $[B_f^K,B_{f'}^K]=0$ and $[C_l^K,C_l^K]=0.$ Also all the boundary plaquette operators and boundary edge operators are projectors with the definition above. 

Similar to a bulk vertex operator, a boundary vertex operator $A_v^K$  imposes a gauge transformation on the group elements on the edges incident at the boundary vertex $v$:
\be\label{Avoperator}
\begin{aligned}
&A_v^K=\sum_{k\in K}A_v^k,\\
&A^k_v\BLvert \fourTetboundary{v}{2}{0}{3}{0}{0}{0.5}{0}\Brangle=\mathcal{A}([0v],[v2],[23],k)\BLvert \fourTetboundary{v'}{2}{0}{3}{0}{0}{0.5}{0}\Brangle,
\end{aligned}
\ee
where the ordering $0<v<2<3$ is assumed, and $0$ is a bulk vertex. 
We relabeled the vertex $v$ by $v'$, so that the new group elements on the edges after acting $A_v^k$ can be expressed by $[0v'],[v'2],[v'3]$ etc., which are defined by the following chain rules:
\be
[0v']\cdot[v'v]=[0v],[v'v]\cdot[v2]=[v'2],[v'v]\cdot[v3]=[v'3],
\ee 
where we set $[v'v]=k$. In Eq. \eqref{Avoperator}, $\mathcal{A}([0v],[v2],[23],k)$ is the amplitude associated with the action of $A_v^k$. The action of $A_v^k$ evolves the original vertex $v$ to a new vertex $v'$, resulting in a new $4$-simplex $0v'v23$ and a new tetrahedron $v'v23$. According to bulk theory in the previous section, we must assign the $4$-simplex $0v'v23$  a $4$-cocycle $[0v'v23]\in H^4[G,U(1)]$. A boundary vertex operator differs from a bulk vertex operator in the newly created temporal tetrahedra made of boundary vertices only, such the $v'v23$ in our example. Hence, for each such new temporal tetrahedron, we should assign a $3$-cochain in $C^3[K,U(1)]$ as a factor in the corresponding amplitude, such as the $3$-cochain $[v'v23]$ for the $v'v23$ in our example. The orientation of such $3$-cochains follow our Convention \ref{conv:TetOrientation}. The amplitude $\mathcal{A}([0v],[v2],[23],k)$ now reads graphically and algebraically as
\be\label{Avampgraph}
\begin{aligned}
\mathcal{A}([0v],[v2],[23],k) &=\Avamplitude{v}{2}{0}{3}{1}{v'}{0.5}
=\cdots [v'v23][0v'v23]\cdots, 
\end{aligned}
\ee
where the $\cdots$ represents those $3$-cochains and $4$-cocycles associated with those tetradedra and $4$-simplexes not shown in the graph. Note that the amplitude of a bulk vertex operator should not contain any $3$-cochains because each bulk tetrahedron is shared by two $4$-simplexes and would contribute two $3$-cochains that cancel out. 

Next, we check whether the bulk vertex and face operators are still commuting projectors in ground states. Consider two boundary vertex operators $A_v^k$ and $A_{v'}^{k'}$. Obviously, if $v$ and $v'$ are the two ends of an edge, $[A_v^k,A_{v'}^{k'}]=0$. What if  $v$ and $v'$ are connected by a boundary edge directly. Without loss of generality, Let us compute an example of such a commutator, i.e., assuming $v(v')=1(2)$ in the following.
\begin{equation}
\label{Avcommutator}
\begin{aligned}
A_{1}^{k}A_{2}^{k'}\BLvert \Avcommutator{1}{3}{2}{4}{1}{5}{0.5}\Brangle=\frac{[12'245][12'24][1'12'35][1'12'3]}{[12'235][12'23][1'12'45][1'12'4]}\BLvert \Avcommutator{1'}{3}{2'}{4}{1}{5}{0.5}\Brangle.\\
A_{2}^{k'}A_{1}^{k}\BLvert \Avcommutator{1}{3}{2}{4}{1}{5}{0.5}\Brangle=\frac{[1'2'245][1'2'24][1'1235][1'123]}{[1'2'235][1'2'23][1'1245][1'124]}\BLvert \Avcommutator{1'}{3}{2'}{4}{1}{5}{0.5}\Brangle.
\end{aligned}
\end{equation}
Then, $A_1^kA_2^{k'}=A_2^{k'}A_1^k$ requires that
\be\label{eq:commequal}
\begin{aligned}
\frac{[12'245][12'24][1'12'35][1'12'3]}{[12'235][12'23][1'12'45][1'12'4]}
=\frac{[1'2'245][1'2'24][1'1235][1'123]}{[1'2'235][1'2'23][1'1245][1'124]}.
\end{aligned}
\ee
We apply the $4$-cocycle condition
\begin{equation*}
\begin{aligned}
\frac{[12'245][1'1245][1'12'25]}{[1'2'245][1'12'45][1'12'24]}=1,
\frac{[12'235][1'1235][1'12'25]}{[1'2'235][1'12'35][1'12'23]}=1,
\end{aligned}
\end{equation*}
to simplify Eq. \eqref{eq:commequal} and obtain the following constraint:
\be\label{eq:generalizedFroC}
\begin{aligned}
[1'12'24]\dd[1'12'24]=[1'12'23]\dd[1'12'23],
\end{aligned}\ee
where all the group variables involved take value in the subgroup $K$. 

Let us temporarily define a function $f:K^{4}\rightarrow U(1)$ by $f(g,h,k,l)=[g,h,k,l]\dd[g,h,k,l]$. Then, Eq. \eqref{eq:generalizedFroC} becomes $f(1'1,12',2'2,24)=f(1'1,12',2'2,23)$. Since the group elements $[23]$ and $[24]$ are arbitrary in $K$. We must have $f(1'1,12',2'2,x)=c$ $\forall x\in K$, where $c$ is a constant. By the normalization condition \eqref{NormalizationCondition}, however, $f(1'1,12',2'2,x=1)=1$; hence, we must have $c=1$. In fact, the other group variables $[1'1],[12'],[2'2]$ in Eq. \eqref{eq:generalizedFroC} are also arbitrary. We can conclude that $f(g,h,k,l)\equiv 1$. Since the example calculation of the commutator $[A^k_v,A^{k'}_{v'}]$ is generic, we would arrive at the same conclusion for the commutator on any two neighbouring boundary vertices $v$ and $v'$. Therefore, requiring the commutativity of neighbouring boundary vertex operators leads to what we call the \textit{generalized Frobenius condition}\footnote{This is a generalization of the usual Frobenius condition\cite{Bullivant2017}}, which can be shown graphically as
\be\label{FroC}
\begin{aligned}
\frobenius{1'}{2'}{3}{1}{1}{2}{0.7}&=1,
\end{aligned}
\ee
or algebraically as
\be\label{generalFroC}
\begin{aligned}
[g,h,k,l]\dd[g,h,k,l]=1,g,h,k,l\in K.
\end{aligned}
\ee
The left hand side of Eq. \eqref{FroC} is a $3$-complex $1'12'23$ due to the commutator $[A_1,A_2]$ and corresponds to the $4$-cocyle $[g,h,k,l]$ in Eq. \eqref{generalFroC} because the group variables involved are arbitrary in $K$. The five tetrahedra composing $1'12'23$ produces five $3$-cochains, whose product corresponds to the $4$-coboundary $\dd[g,h,k,l]$ in Eq. \eqref{generalFroC}. The generalized Frobenius condition \eqref{generalFroC} now can be understood as demanding that if  the four group variables of a $4$-cocycle are restricted in the subgroup $K$, it must be cohomologically trivial. 

Nevertheless, the generalized Frobenius condition \eqref{generalFroC} is an equation that may not have a solution for certain $K\subseteq G$. If we would like our Hamiltonian being exactly solvable, i.e., all operators commuting, the generalized Frobenius condition must hold. If the Hamiltonian is not exactly solvable, we are not sure whether the boundary would be gapped or not. In this paper, we consider exactly solvable Hamiltonians only and thus enforce the generalized Frobenius condition. 

Consequently, given a $4$-cocycle $\omega\in H^4[G,U(1)]$, a subgroup $K\subseteq G$ that allows solutions of the $3$-cochain $\alpha\in C^3[K,U(1)]$ to the generalized Frobenius condition defines a class of physical boundary conditions, each for a solution. In the next section, we shall show that the set of solutions of $\alpha$ actually are in one-to-one correspondence with the $3$-cocyles in $H^3[K,U(1)]$. Therefore, a permissible subgroup $K\subseteq G$ and a $3$-cocycle $\tilde\alpha\in H^3[K,U(1)]$ defines a physical boundary condition for a given bulk theory defined by $G$ and $\omega\in H^4[G,U(1)]$. 

Apart from the proof above, we need also to check the commutator between the boundary and bulk vertex operators and show that $A_v^K$ is a projector. These calculations would not result in any further constraints on our model. We thus leave them to Appendix \ref{appd:vertexOp}. 
\section{One-to-One correspondence between 3-cochain and 3-cocycle}\label{sec:one2one}
We now prove the claim we made in the previous section that given a $K\subseteq G$, the $3$-cochain solutions $\alpha\in C^3[K,U(1)]$ to the generalized Frobenius condition \eqref{generalFroC} are in one-to-one correspondence with the $3$-cocycles in $H^3[K,U(1)]$.
 
\begin{proof}
Let us pick an arbitrary solution $\alpha_0$ of $\alpha$ to Eq. \eqref{generalFroC}. Denote $H^3[K,U(1)]$ explicitly by $\{\tilde\alpha_1, \tilde\alpha_2,\dots,\tilde\alpha_n\}$, where $\tilde\alpha_i$ and $\tilde\alpha_j$ are (representatives of) inequivalent $3$-cocycles. Because $\dd \tilde\alpha_i\equiv 1$ and and $\dd(\alpha\tilde\alpha_i)=\dd \alpha\dd \tilde\alpha_i,\ \forall i$, $H^3[K,U(1)]$ yields a set $\{\alpha_0,\alpha_0\tilde\alpha_1, \alpha_0 \tilde\alpha_2,\dots,\alpha_0\tilde\alpha_n\}$ of solutions to Eq. \eqref{generalFroC}. Here, the operator $\dd$ is the coboundary operator defined in Appendix \ref{app:HnGU1}.

Conversely, consider any solution $\alpha_m$ to Eq. \eqref{generalFroC} other than $\alpha_0$, we have $\omega\dd\alpha_m =\omega\dd\alpha_0=1$. Hence, 
\[ 
\dd\alpha_m=\dd\alpha_0\Rightarrow \dd(\alpha_m\alpha_0^{-1})=1\Rightarrow \alpha_m =\alpha_0\tilde\alpha_m, 
\]
where $\tilde\alpha_m\in H^3[K,U]$ is a $3$-cocycle. It does not matter which solution $\alpha_0$ we choose to generate the set of solutions $\{\alpha_0,\alpha_0\tilde\alpha_1, \alpha_0 \tilde\alpha_2, \dots,\alpha_0\tilde\alpha_n\}\xeq{\mathrm{def}}\{\alpha_i|i=0,\dots, n=|H^3[K,U(1)]|-1\}$. For future convenience, we denote this set of $3$-cochains that specify all possible boundary conditions for a given $K\subseteq G$ by $\Lambda_K$. This establishes the claimed one-to-one correspondence. 
\end{proof}  

Now the question is whether two pairs $(\omega,\alpha)$ and $(\omega',\alpha')$ , where $\omega,\omega'\in H^4[G,U(1)]$ and $\alpha,\alpha'\in \Lambda_K$, give rise to the same topological order with the same boundary condition. We shall report elsewhere the answer to this question. We refer the readers to Ref. \cite{Hu2017c} for the answer to the version of this question for the TQD model of $2+1$-dimensional topological orders with boundaries. 
\section{Ground state wavefunctions on a 3-ball}\label{sec:groundstate}
In this section, we derive the explicit ground-state wavefunctions of the TGT model on a $3$-Ball.
 
 Since here we are  interested in ground states only, we can restrict our concern to the subspace $\Hil^{B_f=1}$ and the boundary degrees of freedom being all in certain subgroup $K\subseteq G$. Moreover, to guarantee the ground states, the bulk $4$-cocycle $\omega$ and the boundary $3$-cochain $\alpha$ must meet the generalized Frobenius condition \eqref{generalFroC}.  

\begin{figure}[h!]
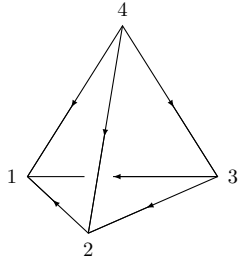

\centering
\fourTet{1}{2}{4}{3}{0}{}{1}{1}
\caption{The simplest triangulation of a $3$-ball.}
\label{fig:3ball1}
\end{figure}

Before deriving the formula of ground state wavefunction on a generic triangulation, let us consider the simplest triangulation of a $3$-Ball Fig. \ref{fig:3ball1}.
All the relevant degrees of freedom are on the boundary edges and thus take value in $K$. Flatness reduces the number of independent degrees of freedom to $3$, and we can choose them to be $h=[12],k=[23],l=[34]$. In this case, the ground-state wavefunction $\Phi_0$ is simply the Dijkgraaf-Witten partition function with all the bulk degrees of freedom integrated out. So, the result is merely the $3$-cocycle associated with the tetrahedron in Fig. \ref{fig:3ball1}. Namely,
\be\label{eq:groundstateball1}
\Phi_{0}=[h,k,l],\ee
where  $[h,k,l]=[1234]$ is the $3$-cocycle associated with the tetrahedron $1234$.

\begin{figure}[h!]
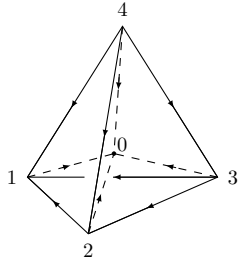

\centering
\fourTet{1}{2}{4}{3}{1}{0}{1}{1}
\caption{Another triangulation of a $3$-ball.}
\label{fig:3ball2}
\end{figure}

Now we consider a more complicated configuration as illustrated in Fig. \ref{fig:3ball2} with a bulk vertex $0$. Instead of using Dijkgraaf-Witten partition function, we take a different strategy to obtain the ground-state wavefunction by finding the function that have an invariant form under the action of $A_v^k$ operators. This method is very effective for the special case where the vertices on the boundary bound a tetrahedron and are all connected to the single bulk vertex directly.  Now the ground-state wavefunction consists of the associated $4$-cocycle and $3$-cochain:
\be\label{eq:groundstateball2}
\Phi_{0}=[ g,h,k,l][h,k,l],  \ee
where $g=[01]$, $h,k,l$ are previously defined, and $[g,h,k,l]=[01234] $ a $4$-cocycle.
  
It's easy to check that the wavefunction \eqref{eq:groundstateball2} is indeed invariant under the action of ground state projector $P^0$. For example, we check how \eqref{eq:groundstateball2}  transforms  under the action of $A_1^x$  
\be
\label{eq:invariantproof}
\begin{aligned}
&A_1^x [g,h,k,l][h,k,l] \ket{\Gamma}\\
=&  \frac{[ g\bar{x},x,h,k][g\bar x,x,hk,l][x,hk,l][x,h,k]}{[g\bar x,x,h,kl][x,h,kl]}[g,h,k,l][h,k,l]\ket{\Gamma}\\
=& [x,h,k,l][g\bar x,xh,k,l]\frac{[x,hk,l][x,h,k][h,k,l]}{[x,h,kl]}\ket{\Gamma}\\
=&  [g\bar x,xh,k,l][xh,k,l] \ket{\Gamma'}\\
=& [g',h',k,l][h',k,l]\ket{\Gamma'},\\
\end{aligned}
\ee
where following $4$-cocycle condition and generalized Frobenius condition are used.
\begin{equation*}
\begin{aligned}
&\frac{[g\bar x,x,h,k][g\bar x,hk,l][g,h,k,l]}{[g\bar x,x,h,kl][x,h,k,l][g\bar x,xh,k,l]}=1;\\
&[x,h,k,l]\frac{[x,hk,l][x,h,k][h,k,l]}{[x,h,kl][xh,k,l]}=1;
\end{aligned}
\end{equation*} 

\begin{figure}[h!]
\centering
\includegraphics[scale=0.1]{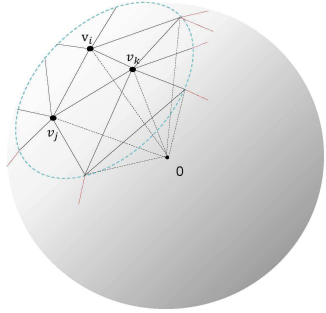}
\caption{A general triangulation of a 3-ball.}\label{fig:ballwf}
\end{figure}

We just discussed examples on two particular triangulations above. In the following we shall derive a generic formula for the ground state wavefunction on a triangulation as in Fig. \ref{fig:ballwf} with one bulk vertex $0$ and $N$ boundary vertices ${v_1,v_2,...v_N}$. We will take the approach of Dijkgraaf-Witten partition functions. Imagine that the graph $\Gamma$ in Fig. \ref{fig:ballwf} evolves from an initial graph $\Gamma'$ where the bulk vertex is $0'$, while the boundary vertices are the same. The ground-state wavefunction $\Phi_0(\Gamma)$ would be identified with the Dijkgraaf-Witten partition function of the evolution from $\Gamma'$ to $\Gamma$, in which the group element $[0'0]$ along the temporal edge $0'0$ is integrated out (i.e., summed over in our discrete case). Hence, $\Phi_0(\Gamma)$ should consist the $4$-cocycles and $3$-cochains associated with all the $3$-simplexes and tetrahedra due to the evolution, with $[0'0]$ summed over. The result is Eq. \eqref{eq:generalspheregfuc}. 
\be\label{eq:generalspheregfuc} 
\begin{aligned}
\Phi_0(\Gamma)=\sum_{\substack{x=[0'0]\in G\\xa_i\in K}}\prod_{\{v_iv_jv_k| 2-\mathrm{simplex}\}}[x,a_i,\bar a_ia_j,\bar a_ja_k]^{-\epsilon(0'0v_iv_jv_k)}[xa_i,\bar a_ia_j,\bar a_ja_k]^{-\epsilon(0'v_iv_jv_k)}.
\end{aligned}
\ee    
 
Here, the product runs over all boundary triangles $v_iv_jv_k$, and $a_i=[0v_i]$. The signs $\epsilon(0'0v_iv_{j}v_{k})$ and $\epsilon(0'v_iv_jv_k)$ follow the conventions defined before. For simplicity, we assume that $0'$ and $0$ are on the opposite sides of the boundary. The negative sign appearing in
front of each $\epsilon$ results from our definition of ground-state wavefunction that $\Phi_0(\Gamma)=\braket{\Gamma}{\Phi_0}$. It is straightforward to prove that $\Phi_0(\Gamma)$ is indeed a $-1$ eigenstate of the Hamiltonian \eqref{eq:HamWithBoundaries}.
If $G=K$ and $4$-cocycle is trivial, the ground-state wavefunction reduces to $\prod_{a_i,a_j,a_k}[a_i,\bar a_ia_j,\bar a_ja_k]^{\epsilon(0v_iv_jv_k)}$.

In the cases where there are more than one bulk vertices, the ground-state wavefunction can be simply obtained from $\Phi_0(\Gamma)$ by performing bulk Pachner moves, which would modify the $4$-cocyles of $\Phi_0(\Gamma)$ only. We refer the reader to Ref. \cite{Wan2014} for the precise algebraic forms of the $3$-dimensional Pachner moves.  
    
\section{GSD on a 3-cylinder}\label{sec:GSD}
\begin{figure}[h!]
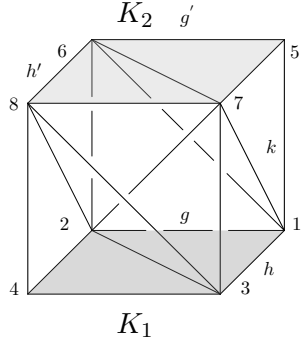

\centering
\torusCube{4}{3}{2}{1}{8}{7}{6}{5}{3}
\caption{A simple triangulation of a $3$-cylinder. Vertex ordering is obvious, and arrows are thus omitted. The surface $5678$ and $1234$ are the two boundaries. We let $[12]=[34]=g, [24]=[13]=h, [56]=[78]=g', [57]=[68]=h', [15]=[26]=[37]=[48]=k$. }\label{fig:cyl}
\end{figure}

On a $3$-ball, the fundamental group of boundary and bulk are both trivial. So the GSD of our model on a $3$-ball is one.
To see how boundary conditions affect the GSD of a $(3+1)$-dimensional topological order, let us consider our model defined on a simple triangulation (see Fig. \ref{fig:cyl}) of a $3$-cylinder (see Fig. \ref{fig:multiboundary}), where the top and the bottom surfaces are the two boundaries with subgroups $K_2, K_1\subseteq G$. 

We let $g\in K_1$ and $g'\in K_2$, but $k\in G$. Since we are interested in ground states only, we let all the triangles in Fig. \ref{fig:cyl} be flat, i.e., we are  working in the subspace $\Hil^{B_f=1}$. Consequently, we have $h'= \bar khk$ and $g'=\bar kgk$.

Note that there are only two boundary vertices in Fig. \ref{fig:cyl}, namely, vertices $1,\ 2,\ 3$, and $4$ are identified, and vertices $5,\ 6,\ 7$, and $8$ are identified. The ground-state projector then reads
\be\label{reduceprojector}
\begin{aligned}
P^0_{3-cyl}=\prod_{v\in\partial\Gamma}A^{K}_v=\frac{1}{|K_1||K_2|}\sum_{x\in K_1}A^x_{1=2=3=4}\sum_{y\in K_2}A^y_{5=6=7=8}.
\end{aligned}\ee Here is an important technical remark. In acting the above operator on the $3$-cylinder, An operator $A^x$ must act on vertices $1,\ 2,\ 3$, and $4$ individually, as if these vertices are different; the identification of $1,\ 2,\ 3$, and $4$ will be automatically accounted for by the periodic boundary condition and there is only one normalization factor $\frac{1}{|K_1|}$. The same procedure applies to $A^y$ on vertices $5,\ 6,\ 7$, and $8$. 
The GSD on a $3$-cylinder is then the trace of $P^0_{3-cyl}$:
\be\label{eq:traceprojector}
\begin{aligned}
GSD_{3-cyl}=\mathrm{Tr}_{\Hil^{B_f=1}}P^0_{3-cyl}
=\sum\limits_{\substack{h,g\in K_{1},\\h',g'\in K_{2},\\k\in G}}\delta_{g',\bar kgk}\delta_{h',\bar khk}\Blangle \smalltorusCube{4}{3}{2}{1}{8}{7}{6}{5}{1} \BRvert P^0_{3-cyl}\BLvert \smalltorusCube{4}{3}{2}{1}{8}{7}{6}{5}{1}\Brangle.
\end{aligned}
\ee
We present the final closed-form of the GSD on a $3$-cylinder as follows but leave the tedious derivation to Appendix \ref{app:GSD3cylinder}. 
\be
\label{eq:3cylinderGSDresult}
\begin{aligned}
GSD_{3-cyl}=\sum\limits_{\substack{x,h,g\in K_{1}\\y,h',g'\in K_{2}}}\sum\limits_{k\in G}\frac{\delta_{g',\bar kgk}\delta_{h',\bar khk}\delta_{g,xg\bar x}\delta_{h,xh\bar x}\delta_{k,xk\bar y}}{|K_{1}||K_{2}|}
\frac{[\bar x,k]_{h\bar g,g}}{[k,\bar y]_{h\bar g,g}}\frac{\eta^{g}(h\bar g,x)}{\eta^{\bar kgk}(\bar kh\bar gk,y)},
\end{aligned}
\ee

where $[a,b]_{m,n}=\frac{[m,a,b]_{n}[a,b,\overline{(ab)}mab]_{n}}{[a,\bar ama,b]_{n}}$ is a doubly-twisted $2$-cocycle, and $\eta^{n}(m,a)$ is defined by $\frac{[\bar a,m]_{n}}{[m,\bar a]_{n}}$ with $[\bar a,m]_{n}$ being a twisted $2$-cochain. The factor $\frac{\eta^{g}(h\bar g,x)}{\eta^{\bar kgk}(\bar kh\bar gk,y)}$ can be rewritten in more compactly as $\frac{1}{\eta^{g,h\bar g}(\bar x,\bar k)}$, which is trivial when the doubly twisted $2$-cocycle is cohomologically trivial. 

This result generalizes the GSD  of the $(2+1)$-dimensional TQD model on a cylinder. It's easy to see that when $G$ is an Abelian group, The  cochain term is simply $1$.
\section{Examples:$G=\Z_2\times \Z_2\times \Z_2$}
Here we consider the TGT model with the gauge group $G=\Z_{2} \times \Z_{2} \times \Z_{2}$ as an example. We check for each subgroup of $G$ the solutions to the generalized Frobenius condition \eqref{generalFroC}. Note that TGT model with this $G$ yields only Abelian excitations\cite{Wan2014}.
The $4$-th cohomology group of $G$ is $\cH^4[ \Z_{2} \times \Z_{2} \times \Z_{2},U(1)] 
=     \Z_{2}^2
         \times \Z_{2}^2 
         \times  \Z_{2}^2
         \times \Z_{2}^2 $.

The generators of the 4-cocycles $\omega$ are (here $(i,j)=(1,2),(2,3)$ or $(1,3)$ and $(i,j,l)=(1,2,3)$):
\begin{align}
&{\omega_{{ \tII}}^{(1st,ij)} (a,b,c,d)=  e^{\big( \frac{\pi \ii  }{ 2   }    (a_i b_j )( c_j +d_j - [c_j+d_j  ]) \big)} },\\
& {\omega_{{ \tII }}^{(2nd,ij)} (a,b,c,d) =  e^{\big( \frac{ \pi \ii  }{ 2  }   (a_j b_i )( c_i +d_i - [c_i+d_i  ])  \big)} },\\
& {\omega_{{\tIII}}^{(1st,ijl)}(a,b,c,d) = e^{\big( \frac{ \pi \ii  }{ 2  }  (a_i b_j )( c_l +d_l - [c_l+d_l  ]) \big)} }, \\
&{\omega_{{\tIII}}^{(2nd,ijl)}(a,b,c,d) = e^{\big( \frac{ \pi \ii  }{ 2  }  (a_l b_i )( c_j +d_j - [c_j+d_j  ]) \big)} },
\end{align}
where $[x]=x\mod 2$.
Any 4-cocycle $\omega$ can be written in terms of the above generators. For example:
\begin{align}
\omega =&\prod_{ \underset{(i,j,l)=(1,2,3)}{(i,j)=(1,2),(2,3),(1,3)} }\omega_{{ \tII}}^{(1st,ij)} \omega_{{ \tII}}^{(2nd,ij)} \omega_{{\tIII}}^{(1st,ijl)} \omega_{{\tIII}}^{(2nd,ijl)} \\
=&\prod_{ \underset{(i,j,l)=(1,2,3)}{(i,j)=(1,2),(2,3),(1,3)} } (\omega_{{ \tII}}^{(1st,ij)})^{p_{{ \tII(ij)}}^{(1st)} } (\omega_{{ \tII}}^{(2nd,ij)})^{p_{{ \tII(ij)}}^{(2nd)} }(\omega_{{\tIII}}^{(1st,ijl)})^{p_{{ \tIII(ijl)}}^{(1st)} } (\omega_{{\tIII}}^{(2nd,ijl)})^{p_{{ \tIII(ijl)}}^{(2nd)} }, 
\end{align}
where ${p_{{ \tII(ij)}}^{(1st)} },\ {p_{{ \tII(ij)}}^{(2nd)} },\ {p_{{ \tIII(ijl)}}^{(1st)} },\ {p_{{ \tIII(ijl)}}^{(2nd)} } \in \Z_{2}$ are called the topological indices\cite{Wang2014c,Wan2014}. Table \ref{tab:Z23sols} then records the $3$-cochain solutions to the generalized Frobenius condition Eq. \eqref{generalFroC} for each generator of the $4$-cocycles and each subgroup $K\subseteq G$. 

\begin{table}[!h]\scriptsize
        \caption{$3$-cochain solutions to the generalized Frobenius condition for $G=\Z_2\times \Z_2\times \Z_2$.  A big $\times$ indicates "no solution".}
        \label{tab:Z23sols}
        \begin{tabular}{|c|c|c|c|c|c|c|c|c|c|}
                \hline
                $K$ & $\omega_0$ & $\omega_{II}^{(1st,12)}$ & $\omega_{II}^{(1st,23)}$ & $\omega_{II}^{(1st,13)}$ & $\omega_{II}^{(2nd,12)}$ & $\omega_{II}^{(2nd,23)}$ & $\omega_{II}^{(2nd,13)}$& $\omega_{III}^{(1st,123)}$&$\omega_{III}^{(2nd,123)}$
                \\
                \hline
                $\{000\}$ & 1 & 1 & 1 &  1 & 1 & 1 & 1  & 1&1
                \\
                \hline
                $\{000,001\}=\Z_2$ & $\Z_2$ & $\Z_2$ & $\Z_2$ & $\Z_2$ & $\Z_2$ & $\Z_2$ &$\Z_2$& $\Z_2$&$\Z_2$
                \\
                \hline
                $\{000,010\}=\Z_2$ & $\Z_2$ & $\Z_2$ & $\Z_2$ & $\Z_2$ & $\Z_2$ & $\Z_2$ &$\Z_2$& $\Z_2$&$\Z_2$
                \\
                \hline
                $\{000,100\}=\Z_2$ & $\Z_2$ & $\Z_2$ & $\Z_2$ & $\Z_2$ & $\Z_2$ & $\Z_2$ &$\Z_2$& $\Z_2$&$\Z_2$
                \\
                \hline
                $\{000,011,001,010\}=\Z_2\times \Z_2$ & $\Z_2^3$ & $\Z_2^3$ & $\times$ & $\Z_2^3$ & $\Z_2^3$ & $\times$ & $\Z_2^3$ & $\Z_2^3$&$\Z_2^3$
                \\
                \hline
                $\{000,101,001,100\}=\Z_2\times 
            \Z_2$ & $\Z_2^3$ & $\Z_2^3$ & $\Z_2^3$ & $\times$ & $\Z_2^3$ & $\Z_2^3$ & $\times$ & $\Z_2^3$&$\Z_2^3$
                \\
                \hline  
                $\{000,110,100,010\}=\Z_2\times \Z_2$ & $\Z_2^3$ & $\times$ & $\Z_2^3$ &$\Z_2^3$ & $\times$ & $\Z_2^3$ & $\Z_2^3$ & $\Z_2^3$&$\Z_2^3$
                \\
                \hline
                $\Z_2\times \Z_2\times \Z_2$ & $\Z_2^7$ & $\times$ & $\times$ & $\times$ & $\times$ & $\times$ & $\times$ & $\times$&$\times$
              
                \\
                \hline
        \end{tabular}
\end{table}
 
\appendix
\section{A brief introduction to cohomology groups $H^n[G,U(1)]$}\label{app:HnGU1}
 We record a few necessary ingredients of the cohomology groups $H^n[G,U(1)]$ of finite groups $G$ in this appendix.

The $n$-th \textit{cochain group} $C^n[G,U(1)]$ of $G$ is an Abelian group of $n$-\textit{cochains}, i.e., functions $c(g_1,\dots,g_n): G^{\times n}\to U(1)$, where $g_i\in G$. The group product reads $c(g_1,\dots,g_n)c'(g_1,\dots,g_n)=(cc')(g_1,\dots,g_n)$. 
The \textit{coboundary operator} $\dd$  maps $C^n$ to $C^{n+1}$, i.e,
\begin{align*}
\dd\ &:C^n\to C^{n+1}\\
&: c(g_1,\dots,g_n)\mapsto (\dd c)(g_0,g_1\dots,g_n),
\end{align*}
where
\begin{align*}
(\dd c)(g_0,g_1\dots,g_n)
= \prod_{i=0}^{n+1}c(\dots,g_{i-2},g_{i-1}g_i,g_{i+1},\dots)^{(-1)^i}.
\end{align*}
The series of variables starts at $g_0$, and ends at $g_{n-1}$. Equation (\ref{4CocycleCondition}) offers the example for $n=4$. 
The nilpotency of $\dd$, i.e., $\dd^2c=1$, leads to the following exact sequence:
\be\label{eq:exactSeq}
\cdots C^{n-1}\stackrel{\dd}{ \to} C^n\stackrel{\dd}{ \to} C^{n+1}\cdots.
\ee 
The images of the coboundary operator, $\mathrm{im}(\dd:C^{n-1}\to C^n)$, form the $n$-th coboundary group, whose elements are dubbed $n$-\textit{coboundaries}. The kernel $\ker(\dd:C^n\to C^{n+1})$ is the $n$-\textit{\textbf{cocycle}} group, whose elements are the $n$-cochains satisfying the \textit{cocycle condition} $\dd c=1$. Equation (\ref{4CocycleCondition}) is the example for $n=4$. The definition of the $n$-th cohomology group is the quotient group that follows from the exact sequence \eqref{eq:exactSeq}:
\[
H^n[G,U(1)]:=\frac{\ker(\dd:C^n\to C^{n+1})}{\mathrm{im}(\dd:C^{n-1}\to C^n)}.
\]
The group $H^n[G,U(1)]$ is Abelian by construction and comprises the equivalence classes of the $n$-cocyles that differ from one another by merely an $n$-coboundary.
A trivial $n$-cocycle is one that can be written as a $n$-coboundary. 

One can define a \emph{slant product} that maps $c_g$ as follows.

\be\label{eq:slantProd}
c_g(g_1,g_2,\dots,g_{n-1})=c(g,g_1,g_2,\dots,g_{n-1})^{(-1)^{n-1}} \prod^{n-1}_{j=1} c(g_1,\dots,g_j,(g_1\cdots g_j)^{-1}g(g_1\cdots g_j),\dots,g_{n-1})^{(-1)^{n-1+j}}.
\ee

The twisted $3$-cocycles and the doubly-twisted $2$-cocycles we have encountered in the main text are examples of this slant product.
\section{Some properties about $A_v^K$ operators}\label{appd:vertexOp}
\begin{figure}[!h]
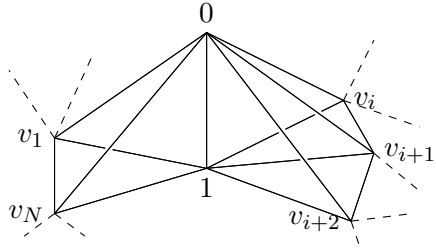

\centering
  \Avprojector
  \caption{A portion of a graph, where some boundary vertices $\{1,v_1,v_2,...,v_N\}$ and a bulk vertex $0$ are shown. We assume that $v_{i+1}>v_i>1$.}
  \label{fig:Avprojector}
\end{figure}

First, let us prove that $A_v^K$is a projector. To this end, we need to show that $A_v^{k'}A_v^k=A_v^{k'k}$. Consider a portion (Fig. \ref{fig:Avprojector}) of a graph $\Gamma$.

We denote the state vector associated with the graph $\Gamma$ by $\ket{\Gamma}$. Without loss of generality, we consider the action of $A^K_v$ on vertex $1$. The action of $A^k_1$ on vertex $1$ would replace vertex $1$ by a new vertex $1'$, with $1'<1$. Then, acting $A^K_v$ twice on vertex $1$ involves terms like $A_{1'}^{k'}A_1^k$ acting on $\ket{\Gamma}$. In the end, $\Gamma$ would become some $\Gamma''$, where vertex $1$ becomes vertex $1''$. We have
\be\label{eq:A1action}
\begin{aligned}
A_{1'}^{k'}A_1^k\ket{\Gamma}&=\delta_{[1'1],k}\delta_{[1''1'],k'}[01'1v_1v_N][1'1v_1v_N][01''1'v_1v_N][1''1'v_{1v_{N}}]\\
& \times\prod_{i=1}^{N}\frac{1}{[01'1v_iv_{i+1}][1'1v_iv_{i+1}][01''1'v_iv_{i+1}][1''1'v_iv_{i+1}]}\ket{\Gamma''},
\end{aligned}
\ee
Using the following instances of the $4$-cocycle condition \eqref{4CocycleCondition} and the generalized Frobenius condition \eqref{generalFroC}
\begin{equation*}
\begin{aligned}
&\frac{[1''1'1v_iv_{i+1}][01''1v_iv_{i+1}][01''1'1v_{i+1}]}{[01'1v_iv_{i+1}][01''1'v_iv_{i+1}][01''1'1v_i]}=1,\\
&\frac{[1''1'1v_1v_{N}][01''1v_1v_{N}][01''1'1v_{N}]}{[01'1v_1v_{N}][01''1'v_1v_{N}][01''1'1v_1]}=1,\\
&[1''1'1v_iv_{i+1}]\frac{[1'1v_iv_{i+1}][1''1'v_{i}v_{i+1}][1''1'1v_i]}{[1''1v_iv_{i+1}][1''1'1v_{i+1}]}=1,\\
&[1''1'1v_1v_{N}]\frac{[1'1v_1v_{N}][1''1'v_{1}v_{N}][1''1'1v_1]}{[1''1v_1v_{N}][1''1'1v_{N}]}=1,
\end{aligned}
\end{equation*}
where $i=1,2...N$, Eq. \eqref{eq:A1action} becomes
\be\label{eq:proofAvprojector}
\begin{aligned}
A_{1'}^{k'}A_1^k\ket{\Gamma}=&\delta_{[1'1],k}\delta_{[1''1'],k'}[01''1v_1v_N][1''1v_1v_N]
\prod_{i=1}^N\frac{1}{[01''1v_iv_{i+1}][1''1v_iv_{i+1}]}
\frac{[01''1'1v_N]}{[01''1'1v_1]}\prod_{i=1}^N\frac{[01''1'1v_i]}{[01''1'1v_{i+1}]}\\
&\times\frac{[1''1'1v_N]}{[1''1'1v_1]}\prod_{i=1}^N\frac{[1''1'1v_i]}{[1''1'1v_{i+1}]}\ket{\Gamma''}\\
&=\delta_{[1'1],k}\delta_{[1''1'],k'}[01''1v_1v_N][1''1v_1v_N]
\prod_{i=1}^N\frac{1}{[01''1v_iv_{i+1}][1''1v_iv_{i+1}]}\ket{\Gamma''}=A_1^{k'k}\ket{\Gamma}.\\
\end{aligned}
\ee
Since $A_v^K$ is an average of $A_v^k$ over $K$, by the rearrangement lemma of group theory, $(A_v^K)^2=A_v^K$. Thus, $A_v^K$ is a projector. One may ask if there are more than one bulk vertices in this graph, would Eq. \eqref{eq:proofAvprojector} still hold? The answer is yes, as from Eq. \eqref{eq:proofAvprojector}, we see that the $4$-cocycle terms and $3$-cochain terms are decoupled, the effect of of $A_v^{k'}A_v^k$ on the tetrahedra and faces around vertex $1$ will always leads to $4$-cocyles and $3$-cochains that can cancel out by applying the $4$-cocycle condition and generalized Frobenius condition, hence arriving at the last equality in Eq. \eqref{eq:proofAvprojector}.

Consider again Fig. \ref{fig:Avprojector}, we can also prove the commutativity of bulk and boundary vertex operators. The vanishing of $[A_0^g,A_1^k]$ requires 
\be\label{brdybulkcommutator}
\begin{aligned}
\frac{[01'1v_iv_N]}{[0'01'v_1v_N]}\prod_{i=1}^N\frac{[0'01'v_iv_j]}{[01'1v_iv_j]}=\frac{[0'1'1v_1v_N]}{[0'01v_1v_N]}\prod_{i=1}^N\frac{[0'01v_iv_j]}{[0'1'1v_iv_j]}.
\end{aligned}
\ee
The $3$-cochains on both sides of the equation cancel and are thus neglected. Using the following instances of the $4$-cocycle condition \eqref{4CocycleCondition}

\begin{equation*}
\begin{aligned}
&\frac{[01'1v_iv_{i+1}][0'01v_iv_{i+1}][0'01'1v_{i+1}]}{[0'1'1v_iv_{i+1}][0'01'v_iv_{i+1}][0'01'1v_i]}=1,
\frac{[01'1v_1v_N][0'01v_1v_N][0'01'1v_N]}{[0'1'1v_1v_{N}][0'01'v_1v_N][0'01'1v_1]}=1,
\end{aligned}
\end{equation*}
Eq. \eqref{brdybulkcommutator} turn out to be
\be\label{identicaleq}
\begin{aligned}
\frac{[0'01'1v_N]}{[0'01'1v_1]}\prod_{i=1}^{N-1}\frac{[0'01'1v_i]}{[0'01'1v_{i+1}]}=1,
\end{aligned}
\ee
which is an identity because all the factors cancel out. Note that when $v'$ and $v$ do not connect directly, $[A_v^g,A_{v'}^k]=0$ is obvious. Together with the proof above, we conclude that bulk and boundary vertex operators commute in the subspace $\Hil^{B_f=1}$. 
\section{The derivation of GSD on a $3$-cylinder}\label{app:GSD3cylinder}

We derive Eq. \eqref{eq:3cylinderGSDresult} in this appendix. We first check how the projector acts on the $3$-cylinder. To simply the calculation, however, we choose to act the vertex operators involved in the ground-state projector \eqref{reduceprojector} on the vertices in Fig. \ref{fig:cyl} in descending order. We have
\be\label{projectoramplitude}
\begin{aligned}
P^0_{3-cyl}&\ket{12,13,15,56,57}=\frac{1}{|K_1||K_2|}\sum_{x\in K_1}\sum_{y\in K_2}\ket{1'2',1'3',1'5',5'6',5'7'}\\
&\times\frac{[2348'8][2678'8]}{[2378'8]}\frac{[678'8][268'8][348'8]}{[248'8][378'8]}\frac{[237'78'][1267'7]}{[1237'7][267'78'][1567'7]}\frac{[137'7][37'78']}{[157'7][567'7][67'78']}\\
&\times\frac{[156'67'][26'67'8']}{[126'67']}\frac{[56'67'][6'67'8'][156'6]}{[26'68'][126'6]}\frac{1}{[15'56'7']}\frac{[15'57']}{[15'56'][5'56'7']}\frac{1}{[234'48']}\frac{[24'48']}{[234'4][34'48']}\\
&\times\frac{[23'34'8'][123'37']}{[23'37'8']}\frac{[123'3][23'34'][3'34'8']}{[13'37'][3'37'8']}\frac{[2'23'7'8'][12'26'7']}{[2'23'4'8'][12'23'7'][2'26'7'8']}\frac{[12'26'][2'26'8']}{[12'23'][2'23'4'][2'24'8']}\\
&\times\frac{[1'15'6'7'][1'12'3'7']}{[1'12'6'7']}\frac{[1'13'7'][1'12'3'][1'15'6']}{[1'15'7'][1'12'6']}.
\end{aligned}
\ee
By renaming of variables, $[12]=[34]=g, [24]=[13]=h, [56]=[78]=g', [57]=[68]=h', [15]=[26]=[37]=[48]=k$, we have
\be\label{projectoramplitudecl}
\begin{aligned}
&P^0_{3-cyl}\ket{g,h,k,g',h'}=\frac{1}{|K_1||K_2|}\sum_{x\in K_1}\sum_{y\in K_2}\ket{xg\bar x,xh\bar x,xk\bar y,yg'\bar y,yh'\bar y}\\
&\times\frac{[ h\bar g,g,k\bar y,y][k,h'\bar{g'},g'\bar y,y][h\bar g,k\bar y,y,g'\bar y][g,k,h'\bar{g'} \bar y,y][k, g'\bar y,y,h'\bar {g'}\bar y][k\bar y,y,h'\bar {g'}\bar y,yg'\bar y]}{[h\bar g,k,g'\bar y,y][g,h\bar g,k\bar y,y][k,h'\bar {g'}\bar y,y,g'\bar y][k,g',h'\bar {g'}\bar y,y][g,k\bar y,y,h'\bar {g'}\bar y][k\bar y,y,g'\bar y,yh'\bar {g'}\bar y]} \\
&\times\frac{[h\bar g\bar x,x,g\bar x,xk\bar y][g,h\bar g\bar x,x,k\bar y][x,h\bar g\bar x,xk\bar y,yg'\bar y][g\bar x,x,k\bar y,yh'\bar {g'}\bar y][x,k\bar y,yg'\bar y,yh'\bar {g'}\bar y][x,g\bar x,xh\bar g\bar x,xk\bar y]}{[h\bar g,g\bar x,x,k\bar y][h\bar g\bar x,x,k\bar y,yg'\bar y][x,h\bar g\bar x,xg\bar x,xk\bar y][g\bar x,x,h\bar g\bar x,xk\bar y][x,k\bar y,yh'\bar {g'}\bar y,yg'\bar y][x,g\bar x,xk\bar y,yh'\bar {g'}\bar y]}\\
&\times\frac{[y,h'\bar {g'}\bar y,yg'\bar y][g'\bar y,y,h'\bar {g'}\bar y][h'\bar {g'},g'\bar y,y ]}{[y,g'\bar y,yh'\bar {g'}\bar y][h'\bar {g'}\bar y,y,g'\bar y][g',h'\bar {g'}\bar y,y]}\frac{[h\bar g\bar x,x,g\bar x][g,h\bar g\bar x,x][x,g\bar x,xh\bar g\bar x]}{[g\bar x,x, h\bar g\bar x][h\bar g,g\bar x,x][x,h\bar g\bar x,xg\bar x]}.
\end{aligned}
\ee
Following Eq. \eqref{eq:traceprojector}, the GSD on a $3$-cylinder then becomes \be\label{eq:3cylinderGSDnosimplify}
\begin{aligned}
 &GSD_{3-cyl}=\mathrm{Tr}P^{0}_{3-cyl}
=\sum\limits_{\substack{x,h,g\in K_{1}\\y,h',g'\in K_{2}}}\sum\limits_{k\in G}\frac{\delta_{g',\bar kgk}\delta_{h',\bar khk}\delta_{g,xg\bar x}\delta_{h,xh\bar x}\delta_{k,xk\bar y}}{|K_{1}||K_{2}|}\\
&\times\frac{[ h\bar g,g,k\bar y,y][k,h'\bar{g'},g'\bar y,y][h\bar g,k\bar y,y,g'\bar y][g,k,h'\bar{g'} \bar y,y][k, g'\bar y,y,h'\bar {g'}\bar y][k\bar y,y,h'\bar {g'}\bar y,yg'\bar y]}{[h\bar g,k,g'\bar y,y][g,h\bar g,k\bar y,y][k,h'\bar {g'}\bar y,y,g'\bar y][k,g',h'\bar {g'}\bar y,y][g,k\bar y,y,h'\bar {g'}\bar y][k\bar y,y,g'\bar y,yh'\bar {g'}\bar y]} \\
&\times\frac{[h\bar g\bar x,x,g\bar x,xk\bar y][g,h\bar g\bar x,x,k\bar y][x,h\bar g\bar x,xk\bar y,yg'\bar y][g\bar x,x,k\bar y,yh'\bar {g'}\bar y][x,k\bar y,yg'\bar y,yh'\bar {g'}\bar y][x,g\bar x,xh\bar g\bar x,xk\bar y]}{[h\bar g,g\bar x,x,k\bar y][h\bar g\bar x,x,k\bar y,yg'\bar y][x,h\bar g\bar x,xg\bar x,xk\bar y][g\bar x,x,h\bar g\bar x,xk\bar y][x,k\bar y,yh'\bar {g'}\bar y,yg'\bar y][x,g\bar x,xk\bar y,yh'\bar {g'}\bar y]}\\
&\times\frac{[y,h'\bar {g'}\bar y,yg'\bar y][g'\bar y,y,h'\bar {g'}\bar y][h'\bar {g'},g'\bar y,y ]}{[y,g'\bar y,yh'\bar {g'}\bar y][h'\bar {g'}\bar y,y,g'\bar y][g',h'\bar {g'}\bar y,y]}\frac{[h\bar g\bar x,x,g\bar x][g,h\bar g\bar x,x][x,g\bar x,xh\bar g\bar x]}{[g\bar x,x, h\bar g\bar x][h\bar g,g\bar x,x][x,h\bar g\bar x,xg\bar x]}.
\end{aligned}
\ee
Using delta functions $\delta_{g',\bar kgk}$, $\delta_{h',\bar khk}$, $\delta_{g,xg\bar x}$, $\delta_{h,xh\bar x}$, $\delta_{k,xk\bar y}$, The $GSD_{3-cyl}$ becomes
\be\label{eq:3cylinderGSD}
\begin{aligned}
 &GSD_{3-cyl}
=\sum\limits_{\substack{x,h,g\in K_{1}\\y,h',g'\in K_{2}}}\sum\limits_{k\in G}\frac{\delta_{g',\bar kgk}\delta_{h',\bar khk}\delta_{g,xg\bar x}\delta_{h,xh\bar x}\delta_{k,xk\bar y}}{|K_{1}||K_{2}|}\\
&\times\frac{[ h\bar g,g,k\bar y,y][k,\bar kh\bar gk,\bar kgk\bar y,y][h\bar g,k\bar y,y,\bar kgk\bar y][g,k,\bar kh\bar gk\bar y,y][k,\bar kgk\bar y,y,\bar kh\bar gk\bar y][k\bar y,y,\bar kh\bar gk\bar y,\bar kgk]}{[h\bar g,k,\bar kgk\bar y,y][g,h\bar g,k\bar y,y][k,\bar kh\bar gk\bar y,y,\bar kgk\bar y][k,\bar kgk,\bar kh\bar gk\bar y,y][g,k\bar y,y,\bar kh\bar gk\bar y][k\bar y,y,\bar kgk\bar y,\bar kh\bar gk]} \\
&\times\frac{[h\bar g\bar x,x,g\bar x,k\bar ][g,h\bar g\bar x,x,k\bar y][x,h\bar g\bar x,k,\bar kgk][g\bar x,x,k\bar y,\bar kh\bar gk][x,k\bar y,\bar kgk,\bar kh\bar gk][x,g\bar x,h\bar g,k]}{[h\bar g,g\bar x,x,k\bar y][h\bar g\bar x,x,k\bar y,\bar kgk][x,h\bar g\bar x,g ,k][g\bar x,x,h\bar g\bar x,k][x,k\bar y,\bar kh\bar gk,\bar kgk][x,g\bar x,k,\bar kh\bar gk]}\\
&\times\frac{[y,\bar kh\bar gk\bar y,\bar kgk][\bar kgk\bar y,\bar kh\bar gk\bar y][\bar kh\bar gk,\bar kgk\bar y,y ]}{[y,\bar kgk\bar y,\bar kh\bar gk][\bar kh\bar gk\bar y,y,\bar kgk\bar y][\bar kgk,\bar kh\bar gk\bar y,y]}\frac{[h\bar g\bar x,x,g\bar x][g,h\bar g\bar x,x][x,g\bar x,h\bar g]}{[g\bar x,x, h\bar g\bar x][h\bar g,g\bar x,x][x,h\bar g\bar x,g]}.
\end{aligned}
\ee
Using the the normalization condition \eqref{NormalizationCondition} and the following instances of the generalized Frobenius condition \eqref{generalFroC}, 
\be\label{frobeniuscond}
\begin{aligned}
&[a,b,\bar x,x]\cdot\dd[a,b,\bar x,x]=1\\
&[b,\bar x,x,a\bar x]\cdot\dd[b,\bar x,x,a\bar x]=1\\
&[x,\bar x,xa\bar x,xb\bar x]\cdot\dd[x,\bar x,xa\bar x,xb\bar x]=1\\
&[\bar x,x,\bar x,xa\bar x]\cdot\dd[x,x,\bar x,xa\bar x]=1, 
\end{aligned}
\ee
the GSD can be rewritten as $GSD_{3-cyl}=\sum\limits_{\substack{x,h,g\in K_{1}\\y,h',g'\in K_{2}}}\sum\limits_{k\in G}\frac{\delta_{g',\bar kgk}\delta_{h',\bar khk}\delta_{g,xg\bar x}\delta_{h,xh\bar x}\delta_{k,xk\bar y}}{|K_{1}||K_{2}|}F(x,h\bar g,g,k)\cdot G(y,h\bar g,k)\cdot\frac{\eta^{g}(h\bar g,x)}{\eta^{\bar kgk}(\bar kh\bar gk,y)}$, where

\begin{equation*}
\begin{aligned}
F(x,h\bar g,g,k)=&\frac{[h\bar g,g,\bar x,x][h\bar g,\bar x,x,g\bar x][x,\bar x,h\bar g,g][\bar x,x,\bar x,g][h\bar g\bar x,x,g\bar x,k][g,h\bar g\bar x,x,\bar xk][x,g\bar x,h\bar g,k]}{[g,h\bar g,\bar x,x][g,\bar x,x ,h\bar g\bar x][x,\bar x,g,h\bar g][\bar x,x,\bar x,h\bar g][g\bar x,x,h\bar g\bar x,k][h\bar g,g\bar x,x,\bar xk][x,h\bar g\bar x,g,k]}\\
&\times\frac{[x,h\bar g\bar x,k,\bar kgk][g\bar x,x, \bar xk,\bar kh\bar gk][x,\bar xk,\bar kgk,\bar kh\bar gk]}{[x,g\bar x,k,\bar kh\bar gk][h\bar g\bar x,x ,\bar xk,\bar kgk][x,\bar xk,\bar kh\bar gk,\bar kgk]},
\end{aligned}\end{equation*}
\begin{equation*}\begin{aligned}
G(y,h\bar g,g,k)=&\frac{[\bar kgk,\bar kh\bar gk,\bar y,y][\bar kgk,\bar y,y,\bar kh\bar gk\bar y][y,\bar y,\bar kgk,\bar kh\bar gk][\bar y,y,\bar y,\bar kh\bar gk][k\bar y,y,\bar kh\bar gk\bar y,\bar kgk]}{[\bar kh\bar gk,\bar kgk,\bar y,y][\bar kh\bar gk,\bar y,y,\bar kgk\bar y][y,\bar y,\bar kh\bar gk,\bar kgk][\bar y,y,\bar y,\bar kgk][k\bar y,y,\bar kgk\bar y,\bar kh\bar gk]}\\
&\times\frac{[k,\bar kgk\bar y,y,\bar kh\bar gk\bar y][k,\bar kh\bar gk,\bar kgk\bar y,y][h\bar g,g,k\bar y,y][g,k,\bar kh\bar gk\bar y,y][h\bar g,k\bar y,y,\bar kgk\bar y]}{[k,\bar kh\bar gk\bar y,y,\bar kgk\bar y][k,\bar kgk,\bar kh\bar gk\bar y,y][g,h\bar g,k\bar y,y][h\bar g,k,\bar kgk\bar y,y][g,k\bar y,y,\bar kh\bar gk\bar y]},
\end{aligned}
\end{equation*}
If we swap $h\bar g$ and $g$ in the definitions above, we find that $F(x,h\bar g,g,k)\rightarrow F(x,h\bar g,g,k)^{-1}$ and $G(y,h\bar g,g,k)\rightarrow G(y,h\bar g,g,k)^{-1}$. This implies that $F(x,h\bar g,g,k)$ and $G(y,h\bar g,g,k)$ have the same mathematical form $[*,*]_{h\bar g,g}$.

We first simplify $F(x,h\bar g,g,k)$ using the  normalization condition \eqref{NormalizationCondition} and the following instances of the $4$-cocycle condition \eqref{4CocycleCondition}. 
\begin{equation*}
\begin{aligned}
&\frac{[x,h\bar g\bar x,k,\bar kgk][g\bar x,x,\bar xk,\bar kh\bar gk]}{[h\bar g\bar x,x ,\bar xk,\bar kgk][x,g\bar x,k,\bar kh\bar gk]}=\frac{[x,h\bar g,\bar xk,\bar kgk][g,x,\bar xk,\bar kh\bar gk][x,h\bar g\bar x,x, \bar xgk][x,g\bar x,x ,\bar xk]}{[h\bar g,x,\bar xk,\bar kgk][x,g,\bar xk,\bar kh\bar gk][x,h\bar g\bar x,x,\bar xk][x,g\bar x,x,\bar xh\bar gk]},\\
&
\frac{[h\bar g\bar x,x,g\bar x,k][x,h\bar g\bar x,x,\bar xgk]}{[x,h\bar g\bar x,g,k]}=\frac{[h\bar g,x,g\bar x,k][x,h\bar g\bar x,x,g\bar x]}{[x,h\bar g,g\bar x,k]},\\
&
\frac{[x,g\bar x,h\bar g,k]}{[g\bar x,x, h\bar g\bar x,k][x,g\bar x,x,\bar xh\bar gk]}=\frac{[x,g,h\bar g\bar x,k]}{[g,x,h\bar g\bar x,k][x,g\bar x,x,h\bar g\bar x]},\\
&
\frac{[x,g,h\bar g\bar x,k]}{[x,h\bar g,g\bar x,k]}=\frac{[g,h\bar g,\bar x,k][x,g,h\bar g,\bar xk][xh\bar g,g,\bar x,k][x,h\bar g,g,\bar xk]}{[xg,h\bar g,\bar x,k][x,g,h\bar g,\bar x][h\bar g,g,\bar x,k][x,h\bar g,g,\bar xk]},\\
&
\frac{[h\bar g,x,g\bar x,k][x,g\bar x,x ,\bar xk]}{[g,x,h\bar g\bar x,k][x,h\bar g\bar x,x,\bar xk]}=\frac{[h\bar g,x,g,\bar xk][g,h\bar g,x,\bar xk][h\bar g,x,g\bar x,x][h\bar g\bar x,g\bar x,x,\bar xk]}{[h\bar g,g,x,\bar xk][g,x,h\bar g,\bar xk][g,x,h\bar g\bar x,x][gx,h\bar g\bar x,x,\bar xk]}.\\
\end{aligned}
\end{equation*}

Note that
\begin{equation*}
\begin{aligned}
[x,\bar xk]_{g,h\bar g}=\frac{[g,h\bar g,x,\bar xk][g,x,\bar xk,\bar kh\bar gk][x,h\bar g,\bar xk,\bar kgk][x,\bar xk,\bar kgk,\bar kh\bar gk][h\bar g,x,g,\bar xk][x,g,h\bar g,\bar xk]}{[h\bar g,g,x,\bar xk][g,x,h\bar g,\bar xk][h\bar g,x,\bar xk,\bar kgk][x,\bar xk,\bar kh\bar gk,\bar kg\bar k][x,h\bar g,g,\bar xk][x,g,\bar xk,\bar kh\bar gk]}.
\end{aligned}
\end{equation*}
We then obtain
\begin{align*}
F(x,h\bar g,g,k)=[x,\bar xk]_{g,h\bar g}&\times\frac{[g,h\bar g\bar x,x,\bar xk][h\bar g,g,\bar x,x][h\bar g,\bar x,x,g\bar x][x,\bar x,h\bar g,g][\bar x,x\bar ,x,g][g,h\bar g,\bar x,k]}{[h\bar g,g\bar x,x,\bar xk][g,h\bar g,\bar x,x][g,\bar x,x,h\bar g\bar x][x,\bar x,g,h\bar g][\bar x,x,\bar x,h\bar g][h\bar g,g,\bar x,k]}\\
&\times\frac{[xh\bar g,g,\bar x,k][x,h\bar g,g,\bar x][x,h\bar g\bar x,x,g\bar x][h\bar g,x,g\bar x,x][xh\bar g,g\bar x,x,\bar xk]}{[xg,h\bar g,\bar x,k][x,g,h\bar g,\bar x][x,g\bar x,x,h\bar gx][g,x,h\bar g\bar x,x][xg,h\bar g\bar x,x,\bar xk]}.
\end{align*}

Keep using relations derived from the $4$-cocylce condition \eqref{4CocycleCondition}, we can simplify $F(x,h\bar g,g,k)$ further. Namely, by
\begin{align*}
&\frac{[h\bar g,g,\bar x,x][g,h\bar g,\bar x,k][g,h\bar g\bar x,x,\bar xk]}{[g,h\bar g,\bar x,x][h\bar g,g,\bar x,k][h\bar g,g\bar x,x, \bar x k]}=\frac{[g,\bar x,x,\bar xk]}{[h\bar g,\bar x,x,\bar x k]},\\
&\frac{[xh\bar g,g,\bar x,k][xh\bar g,g\bar x,x,\bar xk][g,\bar x,x,\bar xk]}{[xg,h\bar g,\bar x,k][xg,h\bar g\bar x,x,\bar xk][h\bar g,\bar x,x,\bar xk]}=\frac{[xh\bar g,g,\bar x,x]}{[xg,h\bar g,\bar x,x]},\\
&\frac{[x,h\bar g,g,\bar x][xh\bar g,g,\bar x,x]}{[x,g,h\bar g,\bar x][xg,h\bar g,\bar x,x]}=\frac{[h\bar g,g,\bar x,x][x,g,h\bar g\bar x,x]}{[g,h\bar g,\bar x,x][x,h\bar g,g\bar x,x]},\\
&\frac{[h\bar g,\bar x,x,g\bar x][\bar x,x,\bar x,g]}{[g,\bar x,x,h\bar g\bar x][\bar x,x,\bar x,h\bar g]}=\frac{[h\bar g\bar x,x,\bar x,g][h\bar g,\bar x,x,\bar x]}{[g\bar x,x,\bar x,h\bar g][g,\bar x,x,\bar x]},\\
&\frac{[h\bar g\bar x,x,\bar x,g][x,h\bar g\bar x,x,g\bar x]}{[g\bar x,x,\bar x,h\bar g][x,g\bar x,x,h\bar g\bar x]}=\frac{[h\bar g,x,\bar x,g][x,g,\bar x,h\bar g][x,h\bar g\bar x,x,\bar x]}{[g,x,\bar x,h\bar g][x,h\bar g,\bar x,g][x,g\bar x,x,\bar x]},\\
&\frac{[h\bar g,x,g\bar x,x][x,g,h\bar g\bar x,x]}{[g,x,h\bar g\bar x,x][x,h\bar g,g\bar x,x]}=\frac{[x,g,\bar x,x][h\bar g,xg,\bar x,x][g,x,h\bar g,\bar x][g,h\bar g,\bar x,x][x,h\bar g,g,\bar x]}{[x,h\bar g,\bar x,x][g,xh\bar g,\bar x,x][h\bar g,x,g,\bar x][h\bar g,g,\bar x,x][x,g,h\bar g,\bar x]},\\
&\frac{[h\bar g,xg,\bar x,x]}{[g,xh\bar g,\bar x,x]}=\frac{[h\bar g,g,x,\bar x][h\bar g,x,\bar x,x]}{[g,h\bar g,x,\bar x][g,x,\bar x,x]},\\
&\frac{[x,g,\bar x,x][x,h\bar g\bar x,x,\bar x][h\bar g,x,\bar x,x]}{[x,h\bar g,\bar x,x][x,g\bar x,x,\bar x][g,x,\bar x,x]}=\frac{[h\bar g\bar x,x,\bar x,x]}{[g\bar x,x,\bar x,x]},\\
&\frac{[h\bar g,\bar x,x,\bar x][h\bar g \bar x,x,\bar x,x]}{[g,\bar x,x,\bar x][g\bar x,x,\bar x,x]}=1,
\end{align*}
we obtain
\begin{equation*}
F(x,h\bar g,g,k)=\eta^{g}(h\bar g,x)\times\frac{[x,\bar xk]_{g,h\bar g}}{[x,\bar x]_{g,h\bar g}}=\eta^{g}(h\bar g,x)[\bar x,k]_{h\bar g,g},
\end{equation*}
where doubly-twisted cocycle condition has been used
\begin{equation*}
\frac{[y,z]_{\bar xwx,\bar xux}[x,yz]_{w,u}}{[xy,z]_{w,u}[x,y]_{w,u}}\bigg|_{wu=uw}=1.
\end{equation*}
Similarly, we can  simplify $G(y,h\bar g,g,k)$. Using the relations
\begin{equation*}
\begin{aligned}
&\frac{[h\bar g,k\bar y,y,\bar kgk\bar y][g,k,\bar kh\bar gk\bar y,y]}{[h\bar g,k,\bar kgk\bar y,y][g,k\bar y,y,\bar kh\bar gk\bar y]}=\frac{[h\bar g,k\bar y,y,\bar kgk][g,k\bar y,\bar kh\bar gk,y][k\bar y,y,\bar kgk\bar y,y][gk\bar y,y,\bar kh\bar gk\bar y,y]}{[h\bar g,k\bar y,\bar kgk,y][g,k\bar y,y,\bar kh\bar gk][h\bar gk\bar y,y,\bar kgk\bar y,y][k\bar y,y,\bar kh\bar gk\bar y,y]},\\
&\frac{[k,\bar kgk\bar y,y,\bar kh\bar gk\bar y][gk\bar y,y,\bar kh\bar gk\bar y,y]}{[k,\bar kgk,\bar kh\bar gk\bar y,y]}=\frac{[\bar kgk\bar y,y,\bar kh\bar gk\bar y,y][k,\bar kgk\bar y,y,\bar kh\bar gk]}{[k,\bar kgk\bar y,\bar kh\bar gk,y]},\\
&\frac{[k,\bar kh\bar gk,\bar kgk\bar y,y]}{[k,\bar kh\bar gk\bar y,y,\bar kgk\bar y][h\bar gk\bar y,y,\bar kgk\bar y,y]}=\frac{[k,\bar kh\bar gk\bar y,\bar kgk,y]}{[\bar kh\bar gk\bar y,y,\bar kgk\bar y,y][k,\bar kh\bar gk\bar y,y,\bar kgk]},\\
&\frac{[k,\bar kh\bar gk\bar y,\bar kgk,y]}{[k,\bar kgk\bar y,\bar kh\bar gk,y]}=\frac{[k\bar y,\bar kh\bar gk,\bar kgk,y][k,\bar y,\bar kh\bar gk,\bar kgk][\bar y,\bar kgk,\bar kh\bar gk,y][k,\bar y,\bar kgk,\bar kh\bar gky]}{[k\bar y,\bar kgk,\bar kh\bar gk,y][k,\bar y,\bar kgk,\bar kh\bar gk][\bar y,\bar kh\bar gk,\bar kgk,y][k,\bar y,\bar kh\bar gk,\bar kgky]},\\
&\frac{[k\bar y,y,\bar kgk\bar y,y][k,\bar kgk\bar y,y,\bar kh\bar gk]}{[k\bar y,y,\bar kh\bar gk\bar y,y][k,\bar kh\bar gk\bar y,y,\bar kgk]}=\frac{[y,\bar kgk\bar y,y,\bar kh\bar gk][k\bar y,\bar kgk,y,\bar kh\bar gk][k\bar y,y,\bar kgk\bar y,\bar kh\bar gky][k\bar y,y,\bar kh\bar gk,\bar kgk]}{[y,\bar kh\bar gk\bar y,y,\bar kgk][k\bar y,\bar kh\bar gk,y,\bar kgk][k\bar y,y,\bar kh\bar gk\bar y,\bar kgky][k\bar y,y,\bar kgk,\bar kh\bar gk]},
\end{aligned}
\end{equation*}
we have
\begin{equation*}
\begin{aligned}
&G(y,h\bar g,g,k)=[k\bar y,y]_{h\bar g,g}\frac{[\bar kgk,\bar kh\bar gk,\bar y,y][\bar kgk,\bar y,y,\bar kh\bar gk\bar y][y,\bar y,\bar kgk,\bar kh\bar gk][\bar y,y,\bar y,\bar kh\bar gk][k\bar y,y,\bar kh\bar gk\bar y,\bar kgk]}{[\bar kh\bar gk,\bar kgk,\bar y,y][\bar kh\bar gk,\bar y,y,\bar kgk\bar y][y,\bar y,\bar kh\bar gk,\bar kgk][\bar y,y,\bar y,\bar kgk][k\bar y,y,\bar kgk\bar y,\bar kh\bar gk]}\\
&\times\frac{[\bar kgk\bar y,y,\bar kh\bar gk\bar y,y][k,\bar y,\bar kh\bar gk,\bar kgk][\bar y,\bar kgk,\bar kh\bar gk,y][k,\bar y,\bar kgk,\bar kh\bar gky][y,\bar kgk\bar y,y,\bar kh\bar gk][k\bar y,y,\bar kgk\bar y,\bar kh\bar gky]}{[\bar kh\bar gk\bar y,y,\bar kgk\bar y,y][k,\bar y,\bar kgk,\bar kh\bar gk][\bar y,\bar kh\bar gk,\bar kgk,y][k,\bar y,\bar kh\bar gk,\bar kgky][y,\bar kh\bar gk\bar y,y,\bar kgk][k\bar y,y,\bar kh\bar gk\bar y,\bar kgky]}.
\end{aligned}\end{equation*}
Rewriting the above again by the following relations
\begin{equation*}\begin{aligned}
&\frac{[y,\bar y,\bar kgk,\bar kh\bar gk][k\bar y,y,\bar kh\bar gk\bar y,\bar kgk][k,\bar y,\bar kh\bar gk,\bar kgk]}{[y,\bar y,\bar kh\bar gk,\bar kgk][k\bar y,y,\bar kgk\bar y,\bar kh\bar gk][k,\bar y,\bar kgk,\bar kh\bar gk]}=\frac{[k\bar y,y,\bar y,\bar kgk]}{[k\bar y,y,\bar y,\bar kh\bar gk]},\\
&\frac{[k,\bar y,\bar kgk,\bar kh\bar gky][k\bar y,y,\bar kgk\bar y,\bar kh\bar gky][k\bar y,y,\bar y,\bar kgk]}{[k,\bar y,\bar kh\bar gk,\bar kgky][k\bar y,y,\bar kh\bar gk\bar y,\bar kgky][k\bar y,y,\bar y,\bar kh\bar gk]}=\frac{[y,\bar y,\bar kgk,\bar kh\bar gky]}{[y,\bar y,\bar kh\bar gk,\bar kgky]},\\
&\frac{[\bar y,\bar kgk,\bar kh\bar gk,y][y,\bar y,\bar kgk,\bar kh\bar gky]}{[\bar y,\bar kh\bar gk,\bar kgk,y][y,\bar y,\bar kh\bar gk,\bar kgky]}=\frac{[y,\bar y,\bar kgk,\bar kh\bar gk][y,\bar kh\bar gk\bar y,\bar kgk,y]}{[y,\bar y,\bar kh\bar gk,\bar kgk][y,\bar kgk\bar y,\bar kh\bar gk,y]},\\
&\frac{[\bar kgk,\bar y,y,\bar kh\bar gk\bar y][\bar kgk\bar y,y,\bar kh\bar gk\bar y,y]}{[\bar kh\bar gk,\bar y,y,\bar kgk\bar y][\bar kh\bar gk\bar y,y,\bar kgk\bar y,y]}=\frac{[\bar y,y,\bar kh\bar gk\bar y,y][\bar kh\bar gk,\bar y,\bar kgk,y][\bar kgk,\bar y,y,\bar kh\bar gk]}{[\bar y,y,\bar kgk\bar y,y][\bar kgk,\bar y,\bar kh\bar gk,y][\bar kh\bar gk,\bar y,y,\bar kgk]},\\
&\frac{[y,\bar kgk\bar y,y,\bar kh\bar gk][y,\bar kh\bar gk\bar y,\bar kgk,y]}{[y,\bar kh\bar gk\bar y,y,\bar kgk][y,\bar kgk\bar y,\bar kh\bar gk,y]}\\&=\frac{[y,\bar y,\bar kgky,\bar kh\bar gk][y,\bar y,\bar kgk,y][\bar y,\bar kh\bar gk,y,\bar kgk][\bar y,\bar kgk,\bar kh\bar gk,y][y,\bar y,\bar kh\bar gk,\bar kgk]}{[y,\bar y,\bar kh\bar gky,\bar kgk][y,\bar y,\bar kh\bar gk,y][\bar y,\bar kgk,y,\bar kh\bar gk][\bar y,\bar kh\bar gk,\bar kgk,y][y,\bar y,\bar kgk,\bar kh\bar gk]},\\
&\frac{[y,\bar y,\bar kgky,\bar kh\bar gk]}{[y,\bar y,\bar kh\bar gky,\bar kgk]}=\frac{[y,\bar y,y,\bar kh\bar gk][\bar y,y,\bar kgk,\bar kh\bar gk]}{[y,\bar y,y,\bar kgk][\bar y,y,\bar kh\bar gk,\bar kgk]},\\
&\frac{[\bar y,y,\bar kh\bar gk\bar y,y][\bar y,y,\bar y,\bar kh\bar gk][y,\bar y,\bar kgk,y]}{[\bar y,y,\bar kgk\bar y,y][\bar y,y,\bar y,\bar kgk][y,\bar y,\bar kh\bar gk,y]}=\frac{[\bar y,y,\bar y,\bar kh\bar gky]}{[\bar y,y,\bar y,\bar kgky]},\\
&\frac{[\bar y,y,\bar y,\bar kh\bar gky][y,\bar y,y,\bar kh\bar gk]}{[\bar y,y,\bar y,\bar kgky][y,\bar y,y,\bar kgk]}=1,
\end{aligned}
\end{equation*}
we arrive at the result
\be
\begin{aligned}
G(y,h\bar g,g,k)=\frac{[k\bar y,y]_{h\bar g,g}}{[\bar y,y]_{\bar kh\bar gk,\bar kgk}}\times\frac{1}{\eta^{\bar kgk}(\bar kh\bar gk,y)}=\frac{1}{[k,\bar y]_{h\bar g,g}}\times\frac{1}{\eta^{\bar kgk}(\bar kh\bar gk,y)}.
\end{aligned}
\ee
These establish the GSD formula Eq. \eqref{eq:3cylinderGSDresult}.

\bibliographystyle{apsrev}
\bibliography{StringNet}

\end{document}